\documentclass[twocolumn]{aastex631}

\usepackage{threeparttable}

\submitjournal{PSJ}

\shorttitle{Classification and characterization of MOID evolution for near-Earth asteroids}
\begin{document}

\title{Classifying and characterizing the evolution of minimum orbit intersection distance for near-Earth asteroids}

\correspondingauthor{Shoucun Hu}
\email{hushoucun@pmo.ac.cn}

\author{Shoucun Hu}
\affil{Purple Mountain Observatory, Chinese Academy of Sciences, Nanjing 210023, China\\}
\affil{CAS Center for Excellence in Comparative Planetology, Hefei 230026, China\\}

\author{Yuanyuan Chen}
\affil{Purple Mountain Observatory, Chinese Academy of Sciences, Nanjing 210023, China\\}
\affil{CAS Center for Excellence in Comparative Planetology, Hefei 230026, China\\}

\author{Xinran Li}
\affil{Purple Mountain Observatory, Chinese Academy of Sciences, Nanjing 210023, China\\}
\affil{CAS Center for Excellence in Comparative Planetology, Hefei 230026, China\\}

\author{Xiuhai Wang}
\affil{Purple Mountain Observatory, Chinese Academy of Sciences, Nanjing 210023, China\\}
\affil{School of Astronomy and Space Science, University of Science and Technology of China, Hefei 230026, China\\}

\author{Xuefeng Wang}
\affil{Purple Mountain Observatory, Chinese Academy of Sciences, Nanjing 210023, China\\}

%% Mark off the abstract in the ``abstract'' environment.
\begin{abstract}
    In this paper, the minimum orbit intersection distances (MOIDs) of near-Earth asteroids (NEAs) over the next 200 years were computed and analyzed in detail. It was shown that the MOID of a NEA relative to the Earth-Moon barycenter (EMB) is usually a superior metric for predicting a potential impact than that relative to the Earth. Subsequently, a novel MOID Evolution Index (MEI) spanning from 0.0 to 9.9 was proposed and the orbits of NEAs are classified into 100 distinct categories by considering the variations of the MOID over time, which is useful for quickly screening and prioritizing hazardous asteroids for future research. Furthermore, it was demonstrated that a linear fitting to the MOID evolution provides a simple yet valid approach for most of the NEAs, which is useful for quickly estimating the MOID value without the need to perform an orbit propagation. As a result, a scheme with several parameters was proposed to characterize the MOID variations as well as the relative position information of the critical points along the orbits associated with the minimum distances. A database incorporating these parameters and the MEI values was therefore established for the cataloged NEAs, enabling the derivation of statistically constrained upper bounds for secular MOID drift rate as function of the semi-major axes. Finally, some special orbital configurations and dynamical mechanisms that may lead to a large deviation from the linear fit or multiple orbit crossings were also investigated, indicating the intricate nature in the patterns of MOID evolution for some NEAs.
\end{abstract}

%% Keywords should appear after the \end{abstract} command.
%% The AAS Journals now uses Unified Astronomy Thesaurus concepts:
%% https://astrothesaurus.org
%% You will be asked to selected these concepts during the submission process
%% but this old "keyword" functionality is maintained in case authors want
%% to include these concepts in their preprints.

\section{Introduction}
Near-Earth asteroids (NEAs) are a class of small bodies that may pose a potential threat to the security of the Earth. Consequently, the assessment of the hazard posed by these objects to the Earth has constituted an important research topic \citep{chesley2002quantifying, chapman2004hazard, milani2006asteroid, farnocchia2015impact, roa2021novel}. The minimum orbit intersection distance (MOID) between two objects orbiting the same primary is defined as the minimum distance between the osculating keplerian orbits \citep{sitarski1968approaches}. While a very small MOID between a NEA and the Earth does not guarantee a collision, a large MOID typically indicates that the asteroid will not pose an imminent threat \citep{tommei2021impact}. In practice, the MOID can be used as an indicator of early impact warnings and as a basis for prioritizing future follow-up observations \citep{milani2005nonlinear, mochi2021new}. Particularly, a NEA with an absolute magnitude of less than 22 (or $\sim$140 m in diameter) is classified as a potentially hazardous asteroid (PHA) if its MOID is smaller than 0.05 au \citep{bowell1994earth}.

As the number of NEAs discovered continues to grow, the necessity for a rapid and precise method to calculate the MOID to assess the potential impact hazard is becoming increasingly apparent \citep{chesley2002asteroid}. Over the decades, many different algorithms have been published in the literature to address this problem, including numerical, analytical, algebraic, and hybrid approaches, as summarized by \cite{hedo2018minimum} and \cite{gronchi2023revisiting}. These methods were also employed to evaluate the maximum number of stationary points of the squared distance function between the orbits \citep{gronchi2002stationary, gronchi2005algebraic}. The issue of uncertainty in the MOID, arising from the inherent ambiguity in the orbital elements of an asteroid, was addressed by \cite{gronchi2006mutual, gronchi2007uncertainty}.

Since the orbits of both the Earth and the asteroids are perturbed by the gravitational forces of other bodies in the solar system (it may also be necessary to consider the non-gravitational effect for some asteroids), the MOID values are actually constantly evolving over time \citep{wisdom1987chaotic, morbidelli1999origin}. For a NEA without having a close encounter with planets, the orbit typically undergoes a secular change in the argument of periapsis ($\omega$) and the longitude of the ascending node ($\Omega$), which is primarily affected by the gravitational perturbation from Jupiter, leading to a secular change in the MOID \citep{opik1951collision}. \cite{gronchi2011secular} and \cite{gronchi2013evolution} investigated the long-term evolution of the MOID using a double-averaged restricted 3-body model and calculated the crossing time using the sampling method of line of variation \citep{milani2005nonlinear, milani2010theory}. Employing the JPL small body integrator, \cite{fuentes2023hazardous} conducted high-fidelity orbital propagation of kilometer-sized near-Earth objects, demonstrating that multiple objects can maintain their MOID within 0.01 au over a 1,000-year timespan.

The current Minor Planet Center (MPC) database only publishes the MOID of NEAs relative to the Earth at a given moment. This is a widely used metric for estimating the potential hazard posed by an asteroid. However, the MOID of some objects may undergo a significant change over a relatively short period of time. It is therefore important to understand and characterize this variation in order to better estimate the future hazard, which may be also helpful to determine which objects should be prioritized for research. The objective of this paper is to propose a new method to classify the orbits of NEAs based on the MOID propagation within a specific period of time, allowing for the capture of the evolution characteristics using a single indicator. Furthermore, this paper is also dedicated to characterizing the MOID evolution using a few parameters, which enables a rapid estimation of the MOID value without the necessity for a time-consuming orbit integration. Both the methodologies are advantageous for elucidating the MOID evolution behavior of a NEA, enabling the construction of a database to record the MOID variation information using several parameters for a NEA.

The following is a description of the organization of this paper. Section 2 provides a detailed description of the model and the dataset utilized to compute the MOID values. In Section 3, a quantitative analysis is performed to demonstrate that the MOID relative to the Earth-Moon barycenter (EMB) is a superior metric for characterizing a potential impact than that relative to the Earth. Section 4 presents the evolution patterns of the MOID for the NEAs and proposes a method to classify the orbits according to the variations. In Section 5, we propose a numerical characterization scheme to interprete the MOID evolution and a statistical analysis is performed. A summary of the results is provided in Section 6.

%%%%%%%%%%%%%%%%%%%%%%%%%%%%%%%%%%%%%%%%%%%%%%%%%%
\section{Model and methodology}
\subsection{Data and model}
According to the NASA/JPL small body database, 35,801 near-Earth asteroids (2,444 of them are PHAs) have been discovered as of September 16, 2024 \footnote[1]{\url{https://ssd.jpl.nasa.gov/tools/sbdb_query.html}}. The initial orbital elements of the asteroids used in this paper were retrieved from the database (9 asteroids that have impacted Earth were excluded). The orbits were propagated using a software package that includes a high-fidelity dynamical model \citep{hu2023peculiar}, which takes into account the gravitational influence of the Sun (including the relativistic effect), the eight planets, the Moon, and the 16 massive main-belt asteroids. The JPL DE441 ephemeris was used to compute the position of the planets and Moon \citep{park2021jpl}. For some NEAs with a substantial number of observations, the non-gravitational effect (such as the Yarkovsky effect) can be solved for by means of orbit determination and was also taken into account in the orbit computation \citep{farnocchia2013near, fenucci2024automated}.

\begin{figure*}
    \centering
    \includegraphics[width=0.6\textwidth]{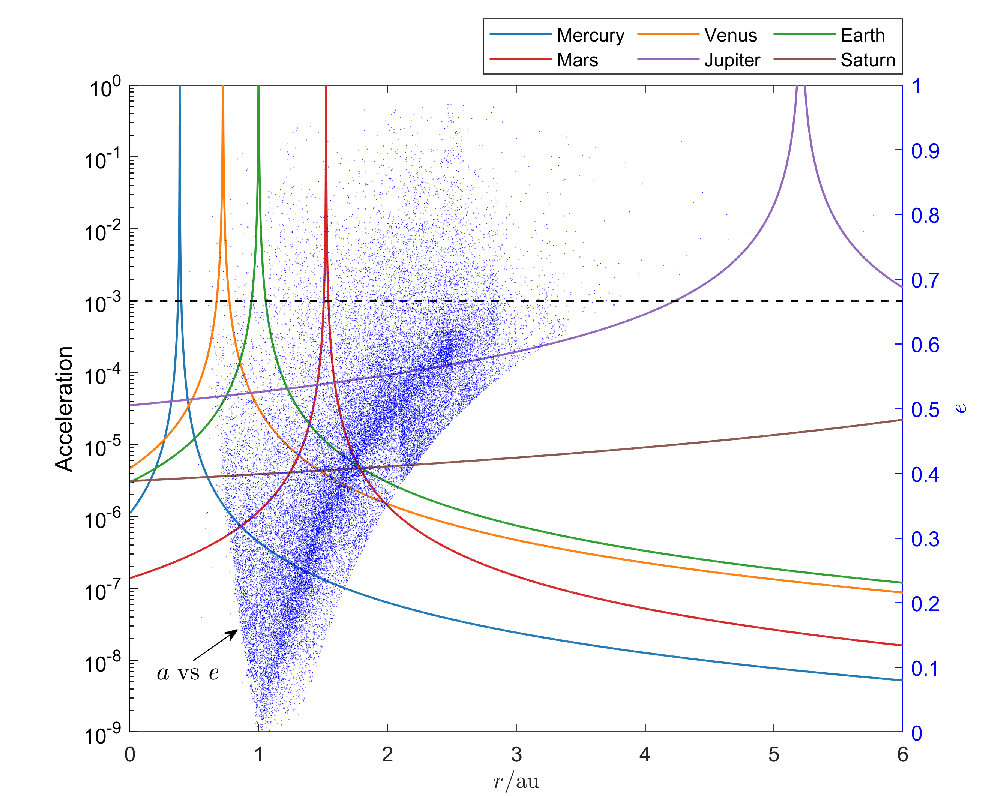}
    \caption{The lines show the variations of gravitational acceleration (normalized by the Sun's gravity) with heliodistance for the six planets. The blue dots illustrate the distribution of semi-major axis versus eccentricity of the NEAs.}
    \label{fig:fig_force_compare}
\end{figure*}

Among the gravitational perturbations exerted by planets, the perturbation by Jupiter is the most prominent one. Fig. \ref{fig:fig_force_compare} presents the relationship between the gravitational acceleration (denoted as $F$, which has been normalized by the Sun's gravity) and the heliodistance for the six planets from Mercury to Saturn, along with the distribution of the semi-major axis versus the eccentricity for the NEA population. It is evident that Jupiter's gravitational influence extends across a substantially broader spatial domain compared to that of the terrestrial planets. When considering the perturbation thresholds of $F > 10^{-3}$ and $F > 10^{-4}$, it is estimated that 6.7\% and 59.6\% of NEAs, respectively, have an aphelion distance that falls within the region of Jupiter's gravitational influence. In contrast, the perturbation exerted by Saturn is not significant, as only a few NEAs are able to approach Saturn's orbit.

\subsection{MOID computation}
Among the six orbital elements of an asteroid, the semi-major axis ($a$), eccentricity ($e$), inclination ($i$), longitude of the ascending node ($\Omega$), and argument of periapsis ($\omega$) can all contribute to the MOID (all the subsequent references to MOID denote values calculated relative to either the Earth or the Earth-Moon barycenter). However, $\Omega$ is usually unimportant due to the low eccentricity of the Earth, as indicated by the simple analytical formula derived by \citep{bonanno2000analytical}
\begin{equation}
    \label{eq:dmin2}
    {d^2} = \frac{{x_0^2{a^2}\left( {1 - {e^2}} \right){{\sin }^2}i}}{{2a - 1 - {a^2}\left( {1 - {e^2}} \right){{\cos }^2}i}}
\end{equation}
in which $d$ is the MOID and 
\begin{equation}
    \label{eq:x0}
    {x_0} = \frac{{a\left( {1 - {e^2}} \right)}}{{1 \pm e\cos \omega }} - 1
\end{equation}
is the nodal distance (measured in au). The `+' in $x_0$ is taken when the critical point is situated in proximity to the ascending node, whereas the `-' denotes that it is in close proximity to the descending node. Note that Eq. \ref{eq:dmin2} is only an approximation to the MOID. It assumes the orbit of the Earth is circular, and the critical point is near the ascending or descending node. Nevertheless, the analytical solution has a very simple and compact expression, which is helpful to gain insight into the evolution of the MOID. For instance, the formula indicates that a low inclination typically results in a comparatively low MOID.

Therefore, in this paper, a numerical approach called the SDG method proposed by \cite{hedo2018minimum, hedo2020minimum} was used to compute the MOID once the orbits were propagated. A medium-term evolution of 200 years (from the epoch $t_0$ of Jan 1, 2025 to the epoch $t_e$ of Jan 1, 2225) is sufficient from an application perspective, as well as the fact that long-term predictions of the orbits for some NEAs are not reliable due to the chaotic nature of their orbits \citep{gronchi2013evolution}.

\section{MOID evolution due to Earth's motion}
The MOID of a near-Earth asteroid is a function of both the asteroid's orbit and the Earth's orbit. It would be advantageous to gain a more comprehensive understanding of the evolution process by quantitatively characterizing the manner in which the MOID evolves when only the Earth's orbit is subject to change. As this topic has not yet been sufficiently addressed in the extant literature, it will be the focus of analysis in this section.

\begin{figure*}
    \centering
    \includegraphics[width=\textwidth]{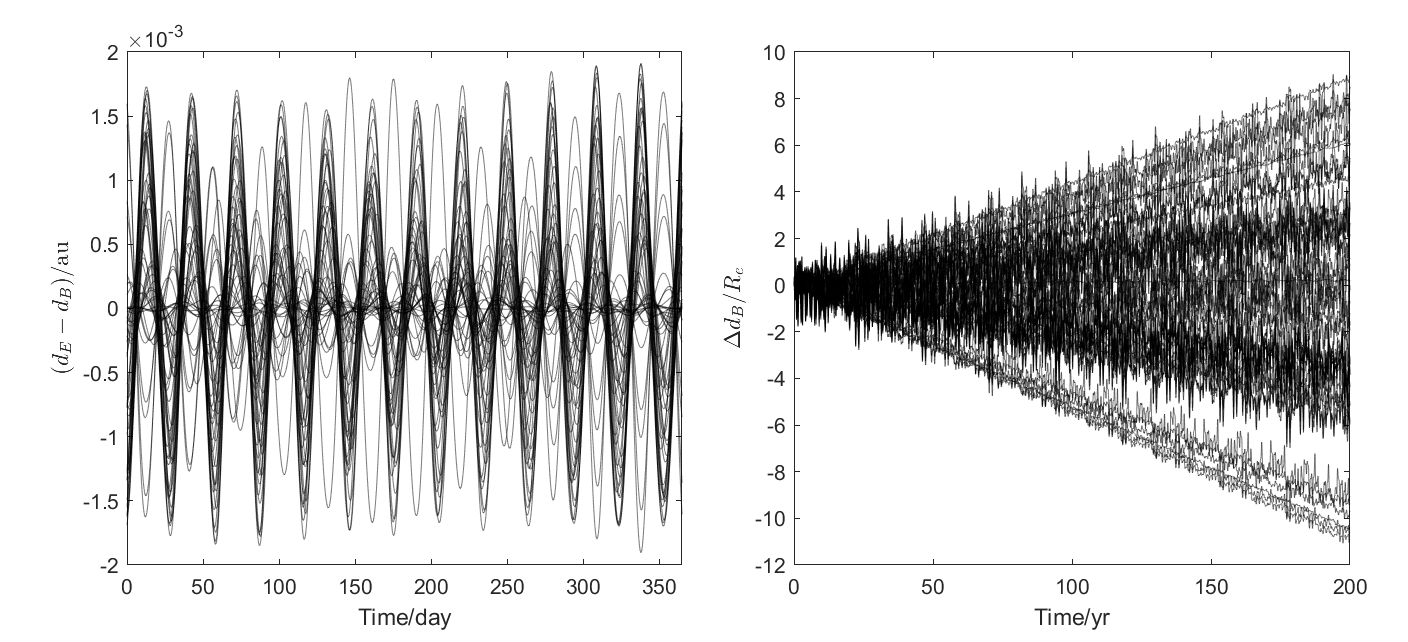}
    \caption{The left panel shows the variations of the difference between $d_E$ and $d_B$ within one year for 50 selected NEAs whose orbits were computed using the two-body model. The right panel depicts the discrepancy in $d_B$ between two models utilized to compute the EMB's orbit for the 50 NEAs: one employs a two-body model, while the other is derived from the DE441 ephemeris.}
    \label{fig:fig_MOIDe_MOIDb}
\end{figure*}

As illustrated in Fig. \ref{fig:fig_force_compare}, Jupiter can exert a perturbation on the Earth's orbit whose magnitude is smaller than 10$^{-4}$ (normalized by the Sun's gravity). Apart from this, however, the Earth's orbit is also subject to perturbations by the Moon and the planets. Here, we define two kinds of MOID, one relative to the Earth's orbit (denoted as $d_E$), and one relative to the Earth-Moon barycenter's (EMB) orbit (denoted as $d_B$). Obviously, the difference of $d_E-d_B$ should only exhibit a periodic change over time. Using the aforementioned dataset, we propagated the orbits of the NEAs for one year (from Jan 1, 2025 to Jan 1, 2026) with a two-body model. Our results indicate that the maximum difference of $d_E-d_B$ is about 0.002 au, which is approximately three quarters of the lunar distance ($LD$, which is about 384,400 km). The variations of the difference for 50 selected NEAs are shown in the left panel of Fig. \ref{fig:fig_MOIDe_MOIDb}.

\begin{figure*}
    \centering
    \includegraphics[width=\textwidth]{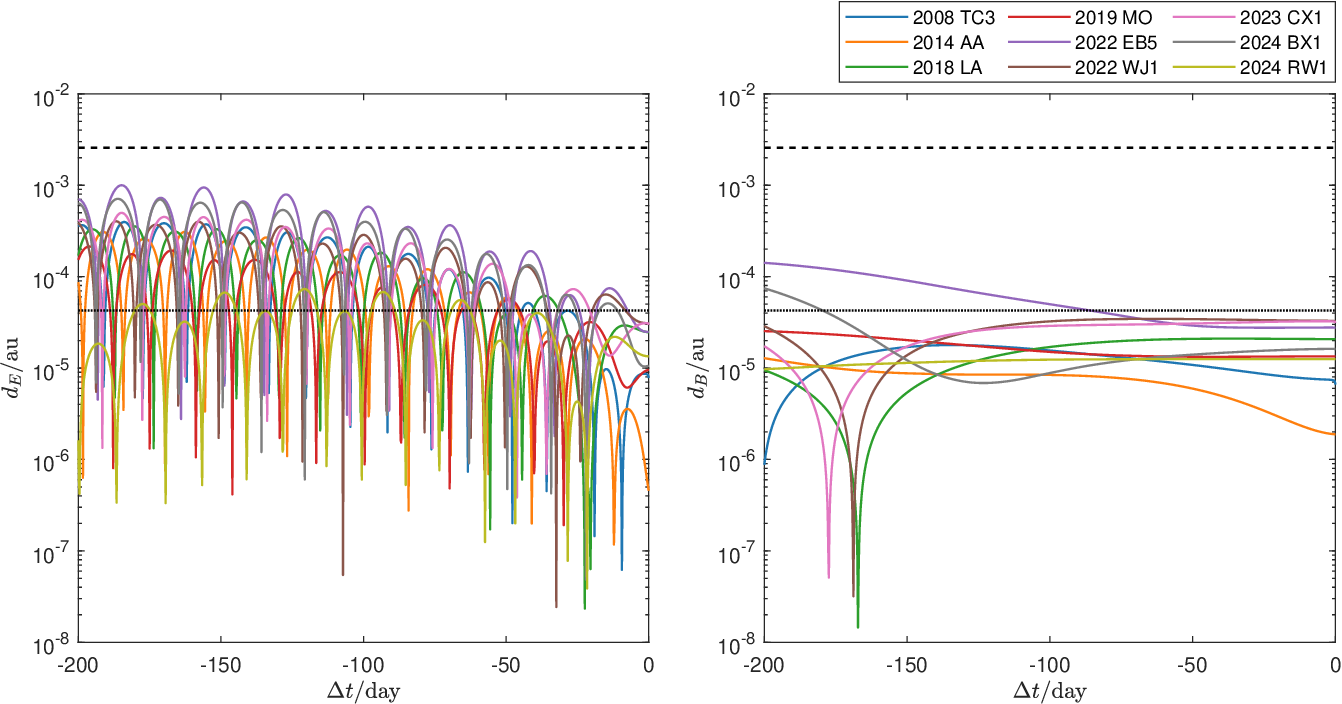}
    \caption{The 200-day backward propagation of $d_E$ (left panel) and $d_B$ (right panel) of nine NEAs that impacted Earth shortly after discovery are illustrated. $\Delta t = 0$ corresponds to the moment of impact. The reference distances of one Earth radius and one lunar distance are indicated by dotted and dashed lines, respectively.}
    \label{fig:fig_moid_bijiao_near_impact}
\end{figure*}

The periodic oscillation of $d_E-d_B$ renders the $d_E$ an unsmooth metric for quantifying the minimum orbit distance between the Earth and an asteroid. To illustrate this point, the orbits of nine near-Earth asteroids that impacted Earth shortly after discovery were integrated backwards for 200 days. This allows for an investigation of the pre-impact evolution of both the two kinds of MOID, as shown in Fig. \ref{fig:fig_moid_bijiao_near_impact}. It is evident that $d_B$ evolves more smoothly than $d_E$. The majority of the orbits have a $d_B$ value smaller than $4\times10^{-5}$ au (within one Earth radius) during the 200 days prior to the impact. In contrast, $d_E$ varies frequently, with the maximum magnitude reaching up to $1\times10^{-3}$ au within the same period. This phenomenon can be attributed to the fact that, during the 200-day pre-impact period, asteroids are generally unlikely to encounter significant gravitational interactions with other terrestrial planets, and Jupiter's gravitational influence remains unimportant due to the relatively large distance. As a result, their orbital elements exhibit smooth, gradual variations. This behavior underscores the effectiveness of the $d_B$ metric in characterizing a potential impact, though this advantage may diminish when considering longer timescales, where additional dynamical effects could come into play. Consequently, we propose to use the $d_B$ rather than the $d_E$ to characterize the minimum distance, though this difference is unimportant for high-MOID cases.

In addition to the short-period motion resulting from the Moon, the Earth's orbit is also subject to the gravitational perturbations by other planets, especially Jupiter. This effect can also be well represented using the EMB's orbit. Consequently, similar to the previous method, we compute the MOID $d_B$ using two different models: one utilizing the EMB's orbit computed with the two-body model and the other derived from the DE441 ephemeris. Our results demonstrate that the discrepancy between the models exhibits a secular change in the MOID of up to 12 $R_e$/200 years, or 0.06 $R_e$/yr ($R_e$ is the Earth radius). This effect is relatively small in comparison to the influence of the short-period motion around the EMB. For any NEA with a perturbed orbit, a variation in the MOID exceeding this value should be attributed to the change of the asteroid's orbit.

Considering the above effects, the minimum distance between the orbits of the Earth and a NEA is approximately quantified using $d_B$ (hereafter simply referred to as $d$) instead of $d_E$ in the following analysis. This approach helps to remove unnecessary periodic variations while preserving the long-term effect. In addition to the orbit change of the EMB, the alteration in the orbit of a NEA typically plays a more significant role in determining the minimum distance.

\section{MOID evolution and classification}
\subsection{Patterns of MOID evolution}
\begin{figure*}
    \centering
    \includegraphics[width=\textwidth]{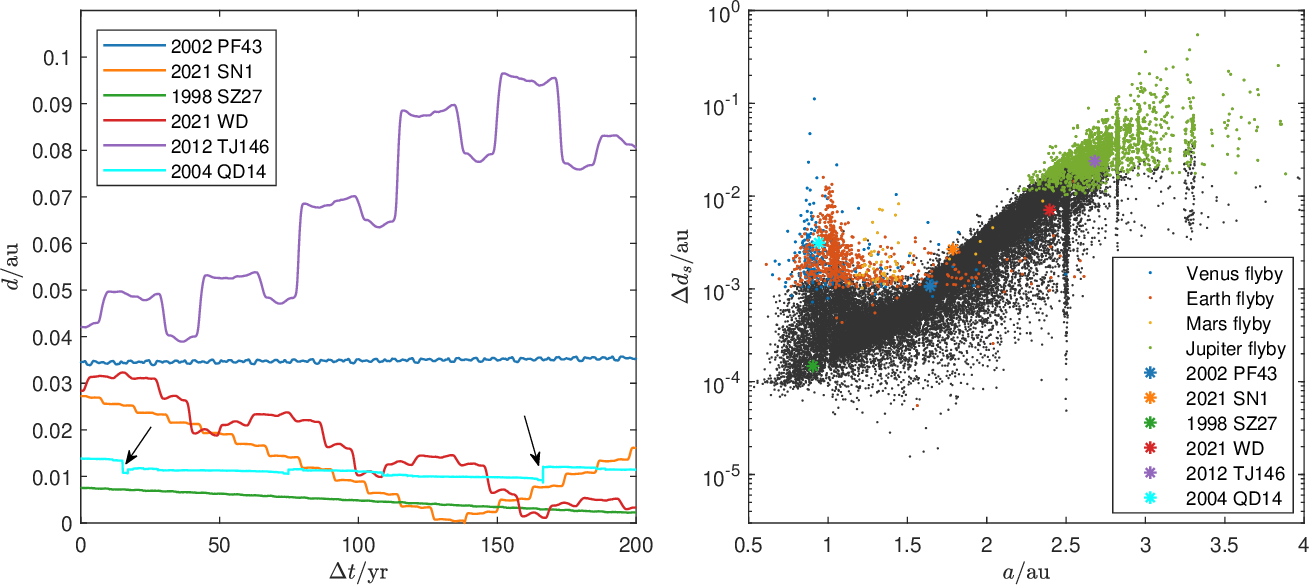}
    \caption{The left panel shows the 200-year MOID evolution of 6 selected PHAs. $\Delta t=t-t_0$ is the time elapsed since the initial epoch $t_0$ (Jan 1, 2025). The right panel depicts the maximal amplitude of short-period variation of the MOID as a function of the semi-major axis. The objects that have close encounters with Venus, Earth, Mars and Jupiter are marked in different colors. The 6 PHAs in the left panel are also labeled (with asterisks) in the right panel.}
    \label{fig:fig_MOID_pattern}
\end{figure*}

To demonstrate the typical evolution patterns of the MOID over the 200-year timespan, we selected 6 PHAs from the database. The MOID evolution for the objects is presented in the left panel of Fig. \ref{fig:fig_MOID_pattern} labeled with different colors. The results reveal significant diversity in the variation patterns among the selected asteroids. Of particular interest is the observation that some of these objects exhibit periodic oscillations superimposed on a linear trend, as also observed by \cite{valsecchi2003resonant}, \cite{gronchi2011secular} and \cite{gronchi2013evolution}. This can be explained by Jupiter's gravitational perturbation which can lead to a secular precession in the argument of perihelion, driving long-period oscillations in the MOID, while the evolution manifests as a quasi-linear trend over short-term timescales of several hundred years \citep{murray1999solar}. For example, the MOIDs of 1998 SZ27, 2002 PF43, and 2021 WD evolve almost linearly in time, with their oscillation amplitudes following a progressive enhancement pattern. This diversity suggests varying magnitudes of gravitational perturbation from planetary interactions. Moreover, the secular drift rate differs among these objects: 2021 WD undergoes the most rapid one, while 2002 PF43 preserves exceptional invariability throughout the interval. For 2021 SN1, it exhibits an inflection point at $\Delta t$ = 138 yr, attributed to the asteroid-EMB orbital intersection. For 2004 QD14, two abrupt changes are observed at $\Delta t$ = 15 yr and 166 yr due to Venus flybys. For 2012 TJ146, however, the variations manifest significant variations because its large aphelion (4.4 au) places it under strong Jupiter gravitational influence. For these objects, the intricate variations cannot be adequately represented by a linear model. A comprehensive analysis of the orbital configuration evolution and gravitational perturbation mechanisms underlying the observed deviation from linearity will be presented in Section \ref{section5}.

\subsection{Short-period oscillations}
The periodic oscillations observed in the MOID propagation are predominantly attributable to the short-period variations in the orbital elements of asteroids. To obtain a better comprehension, the maximal amplitude of these short-period oscillations (denoted as $\Delta d_s$), defined as the maximal difference between the maximum and minimum values within a single asteroid's orbital period over the 200-year timespan, is calculated. The results for the NEAs are illustrated in the right panel of Fig. \ref{fig:fig_MOID_pattern} in relation to the semi-major axis.

The results reveal that the maximum amplitude spans over four orders of magnitude across the population. The semi-major axis dependency manifests in two distinct regimes: For $a>1.5$ au (i.e., beyond the orbital semi-major axis of Mars), the observed upper limit shows a positive correlation with $a$, primarily attributed to Jupiter's overwhelming gravitational dominance. When $a$ surpasses 2.2 au, orbital perturbations become sufficiently strong to induce MOID variations exceeding 0.01 au per orbital cycle. Conversely, for $a < 1.5$ au, dynamical complexity increases significantly due to possible close encounters with terrestrial planets (Earth, Venus, and Mars). Notably, while terrestrial planet flybys may occur at larger $a$ values, their orbital perturbations may remain secondary to Jupiter's dominant gravitational influence.

To quantitatively understand the impact of close encounters between NEAs and major planets (recorded as a close encounter when the distance is less than five times the Hill radius) on the MOID, we calculated the maximum variation amplitude of the MOID (denoted as $\Delta d_f$, defined as the maximum value minus the minimum value) for these asteroids within the orbital segment where the planetary distance is less than five times the Hill radius. By setting a critical value $\Delta d_{fc}$ ranging from 0.001 to 0.05 au, we computed the percentage of asteroids with $\Delta d > \Delta d_{fc}$ relative to the total number of NEAs, as shown in Tab. \ref{tab:closeflyby}. The results show that 13.27\% of NEAs showing variations exceeding 0.001 au for Jupiter's flyby, compared to 1.98\% and 0.71\% for Earth and Venus respectively. The percentages decrease significantly above the 0.01 au threshold, with 5.88\%, 0.03\% and 0.03\% for these planets respectively. Notably, the vast majority (83.9\%) of NEAs exhibit either no close planetary encounters or encounters that induce negligible perturbations to their MOID, with variations smaller than 0.001 au. The right panel of Fig. \ref{fig:fig_MOID_pattern} highlights objects where encounters with Venus, Earth, and Mars result in MOID variations exceeding 0.001 au, while encounters with Jupiter produce variations greater than 0.01 au.

\begin{table*}
    \centering
    \caption{Percentages of NEAs that satisfy $\Delta d_f > \Delta d_{fc}$ after flybys with different planets.}
      \begin{tabular}{ccccccccccccccc}
      \toprule
      $\Delta d_{fc}$/au & 0.001 & 0.002 & 0.003 & 0.004 & 0.005 & 0.006 & 0.007 & 0.008 & 0.009 & 0.01 & 0.02 & 0.03 & 0.04 & 0.05 \\
      \hline
      Mercury & 0.00 & 0.00 & 0.00 & 0.00 & 0.00 & 0.00 & 0.00 & 0.00 & 0.00 & 0.00 & 0.00 & 0.00 & 0.00 & 0.00 \\
      Venus & 0.71 & 0.32 & 0.16 & 0.10 & 0.07 & 0.06 & 0.05 & 0.04 & 0.03 & 0.03 & 0.01 & 0.01 & 0.01 & 0.00 \\
      Earth & 1.98 & 0.74 & 0.34 & 0.17 & 0.10 & 0.07 & 0.05 & 0.04 & 0.03 & 0.03 & 0.00 & 0.00 & 0.00 & 0.00 \\
      Mars & 0.15 & 0.05 & 0.03 & 0.01 & 0.01 & 0.01 & 0.01 & 0.01 & 0.00 & 0.00 & 0.00 & 0.00 & 0.00 & 0.00 \\
      Jupiter & 13.27 & 11.80 & 10.57 & 9.60 & 8.74 & 8.06 & 7.42 & 6.82 & 6.34 & 5.88 & 2.92 & 1.56 & 0.96 & 0.59 \\
      Saturn & 0.02 & 0.02 & 0.01 & 0.01 & 0.01 & 0.01 & 0.01 & 0.01 & 0.01 & 0.00 & 0.00 & 0.00 & 0.00 & 0.00 \\
      \hline
      Total & 16.14 & 12.93 & 11.10 & 9.89 & 8.93 & 8.20 & 7.53 & 6.91 & 6.42 & 5.94 & 2.94 & 1.57 & 0.96 & 0.59 \\
      \hline
      \end{tabular}
    \label{tab:closeflyby}
\end{table*}

Additionally, as observed in the right panel of Fig. \ref{fig:fig_MOID_pattern}, a significant proportion of objects display perturbation amplitudes well below the other objects with the similar semi-major axis. This attenuation can be attributed to some low-inclination or high-inclination objects. For low-inclination objects, they experience weaker Jupiter-induced torque perturbations, which can effectively stabilize the inclination evolution and constrain the MOID variations. Moreover, as observed in Eq. \ref{eq:dmin2}, $\sin i$ appears in the numerator, which can significantly reduce the MOID variation even the semi-major axis, eccentricity, or other parameters are perturbed. On the other hand, high-inclination objects may maintain larger angular separations from the ecliptic plane, reducing the gravitational interactions with major planets and suppressing the MOID variations. The results indicate that 62\% of NEAs exhibit $\Delta d_s$ values smaller than the maximum MOID variation amplitude of 0.002 au induced by the Moon, underscoring the necessity of accounting for and removing these periodic oscillations in our MOID calculations.

\subsection{Classification}
\label{sec:Classification}
The above results demonstrate significant diversity in MOID propagation characteristics among near-Earth asteroids. Such variability highlights the need for a simplified diagnostic indicator to efficiently characterize MOID evolution for individual NEAs. Inspired by the method of quantifying the orbital uncertainty of an asteroid using a condition code from 0 to 9 (as used in the JPL's Small-Body Database), we propose a novel two-digit metric termed MOID Evolution Index (MEI) to systematically quantify the MOID variation according to the minimum ($d_1$) and maximum ($d_2$) MOID values over the 200-year timescale. The MEI is formulated as a decimal value $m.n$, where:

\begin{itemize}
    \item $m$ (decimal digit, 0-9) represents the level of the minimal MOID that can be attained within the 200-year period. Smaller $ m $ values may statistically correspond to higher potential close-approach risks. $m$ is defined as:
$$
    \left\{ \begin{array}{l}
    m = {\rm{INT}}\left[ {{{\log }_2}\left( {\frac{{{d_1}}}{{0.05\;{\rm{au}}}}} \right)} \right] + 6\\
    m = 0\;{\rm{if}}\;m < 0;\;\;\;m = 9\;{\rm{if}}\;m > 9
    \end{array} \right.
$$
in which $\rm{INT}$ is a function that returns the largest integer smaller than the argument.
    \item $n$ (decimal digit, 0-9) quantifies the maximal range of MOID variations over the same interval, with higher $ n $ values reflecting greater orbital instability or more rapid MOID drift. $n$ is defined as:
$$
    \left\{ \begin{array}{l}
    n = {\rm{INT}}\left[ {{{\log }_2}\left( {\frac{{{d_2} - {d_1}}}{{0.00256\;{\rm{au}}}}} \right)} \right] + 3\\
    n = 0\;{\rm{if}}\;n < 0;\;\;\;n = 9\;{\rm{if}}\;n > 9
    \end{array} \right.
$$
\end{itemize}

\begin{figure*}
    \centering
    \includegraphics[width=0.8\textwidth]{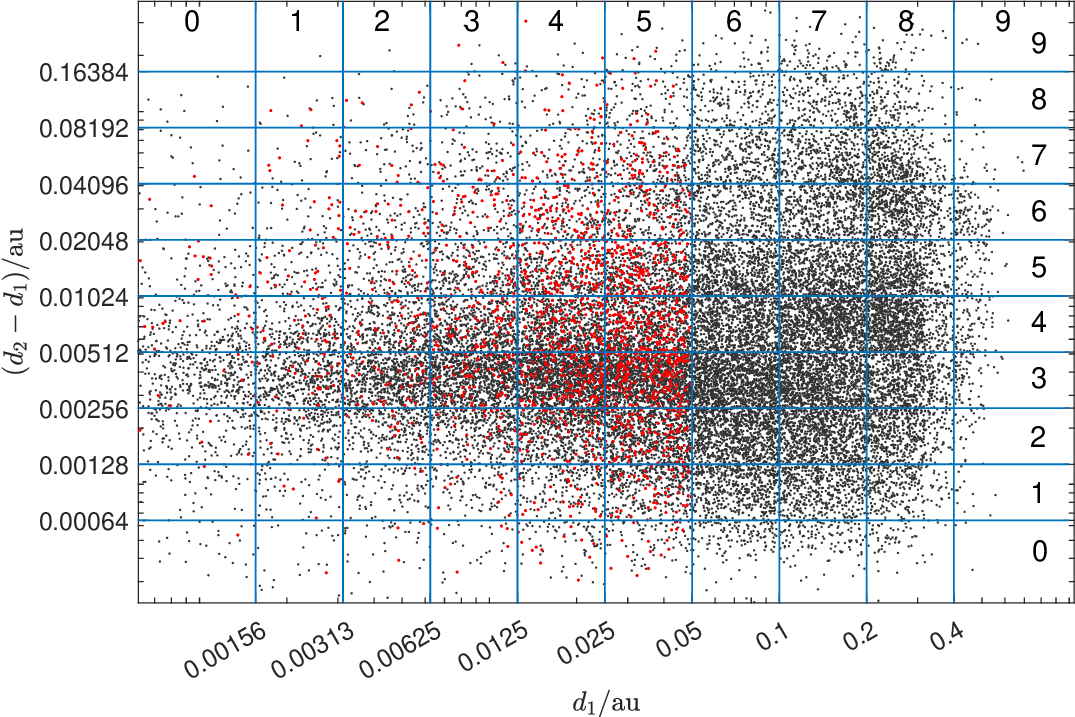}
    \caption{Illustration of the MOID evolution classification scheme using the two-digit MEI metric $m.n$, with the $m$-value defined based on $d_1$ and $n$-value on $d_2$-$d_1$. The cataloged NEAs are classified and plotted in black within their respective grid cells, with the PHAs distinctly highlighted in red.}
    \label{fig:fig_MEI_classify}
\end{figure*}

The MEI spans a discrete scale from 0.0 to 9.9, enabling concise yet information-rich comparisons of asteroid-Earth encounter profiles. For instance, an asteroid with MEI = 1.3 would exhibit a low MOID minimum (high hazard potential) but limited orbital variability, whereas MEI = 5.8 would indicate a moderate MOID minimum coupled with significant dynamical evolution. This dual-parameter scheme provides researchers and planetary defense stakeholders with an intuitive framework to prioritize objects requiring detailed risk analysis or long-term monitoring. Additionally, the above definition is intentionally designed to ensure that PHAs satisfy $m \le 5$, with $ n \le 2$ assigned to objects whose maximum MOID variations remain within one lunar distance (384,400 km $\approx$ 0.00256 au) and $ n \le 4$ applied to those with variations confined to approximately one Earth Hill radius (0.01 au). Fig. \ref{fig:fig_MEI_classify} demonstrates the 10$\times$10 classification grid system, which categorizes all cataloged NEAs into distinct classes based on their MEI values. The PHAs, represented by red dots, exhibit a wide distribution across both $m$- and $n$-values, highlighting the granularity and discriminative capability of the MEI classification system.

The MEI metric assigns the two digits $m$ and $n$ to prioritize NEA risk assessment and it is well defined regardless of the complexity of the MOID variations. This design reflects two key principles: (1) orbits achieving lower minimum MOID values (smaller $m$) are deemed higher-risk due to their closer potential encounters with Earth, and (2) for asteroids sharing the same $m$-value, a larger $n$-value indicates greater orbital configuration changes, which statistically reduces the low-MOID dwell time and, consequently, the probability of close approaches.

However, the comparative risk assessment based on this metric faces inherent limitations when the primary digits ($m$) differ. For example, an asteroid with MEI = 1.0 does not necessarily pose a lower risk than one with MEI = 0.9. The former may exhibit prolonged orbital proximity to Earth, while the latter, despite achieving closer Earth approaches, likely represents only a transient encounter. Furthermore, the metric fails to capture the detailed time-dependent dynamics of orbital evolution. Notably, it does not incorporate critical factors such as actual geocentric distance variations during close approaches or physical properties like asteroid size, both of which are essential for comprehensive hazard evaluation. Consequently, while the MEI effectively categorizes MOID variations and serves as a valuable preliminary screening tool, its ability to holistically assess long-term risks necessitates refinement through supplementary dynamical analysis and physical characterization.

\section{MOID evolution characterization}
\label{section5}
\subsection{Linear fitting}
While the MEI metric effectively captures the minimum MOID value and its maximum variation range over the 200 years, it inherently lacks the capacity to estimate values at arbitrary intermediate epochs. This limitation motivates our investigation into alternative characterization methods. As exemplified by asteroids (523950) 1998 SZ27, (612380) 2002 PF43, and 2021 WD in Fig. \ref{fig:fig_MOID_pattern} (left panel), these objects exhibit near-linear secular trends in their MOID evolution. The observed notable linearity strongly suggests that linear approximations could effectively capture their essential evolutionary characteristics.

\begin{figure*}
    \centering
    \includegraphics[width=\textwidth]{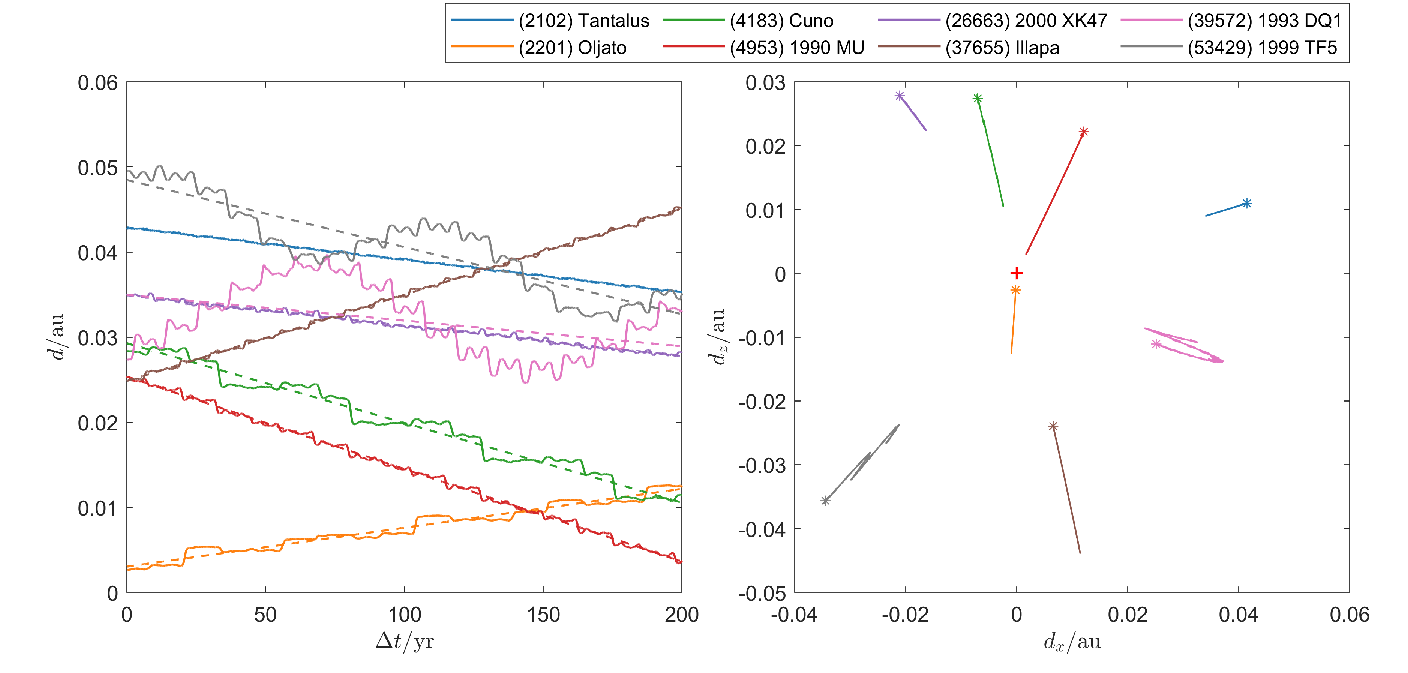}
    \caption{The left panel depicts the fitting results of the MOID variations for eight selected NEAs. The solid lines represent the precise values, while the dashed lines illustrate the linear fits. The right panel depicts the corresponding migration of the critical points projected on the $x-z$ plane in the migration frame. The asterisks indicate the initial position of the points, and the red plus symbol represents the origin of the frame (which is the position of the critical point related to the EMB's orbit).}
    \label{fig:fig_fit_good}
\end{figure*}

Using the least-squares criterion, the MOID at epoch $t$ can be statistically represented through linear fitting
\begin{equation}
    \label{eq:fit0}
    d\left( t \right) = {d_0} + k \cdot \left( {t - {t_0}} \right) + \Delta d\left( t \right), \;\;\; t \in [t_0, t_e]
\end{equation}
where $k$ quantifies the fitted secular drift rate (au/yr or $R_e$/yr), $d_0$ is the initial offset at $t_0$, and $\Delta d(t)$ denotes residual deviations from the linear trend. The optimal parameters $d_0$ and $k$ are determined by minimizing the sum of squared residuals:
\begin{equation}
    \label{eq:fit00}
    \min {\sum\limits_{i = 0}^N {\left[ {{d_0} + k \cdot \left( {{t_i} - {t_0}} \right) - d(t_i)} \right]} ^2}
\end{equation}
where $t_i - t_{i-1}$ = 1 day, $i=1,...,N$, with $t_N = t_e$ and $d(t_i)$ is the high-fidelity MOID at $t_i$ derived from the numerical orbital integrations. In addition, the maximum absolute residual
\begin{equation}
    \label{eq:dmax}
    \varepsilon = \max \left| {{d_0} + k \cdot \left( {{t_i} - {t_0}} \right) - d(t_i)} \right|,\;\;\;i = 0,\;...,\;N
\end{equation}
is used to quantify the upper limit of the uncertainty in the fitting. To illustrate this, the left panel of Fig. \ref{fig:fig_fit_good} presents the fitting results for eight selected NEAs. In the plot, solid lines represent the precise values of $d$, while dashed lines depict the corresponding linear fits. However, for certain asteroids that encounter highly complex dynamic environments, the  MOID evolution may exhibit a similarly intricate and non-linear behavior. This phenomenon will be explored in detail in the following sections.

\begin{figure*}
    \centering
    \includegraphics[width=\textwidth]{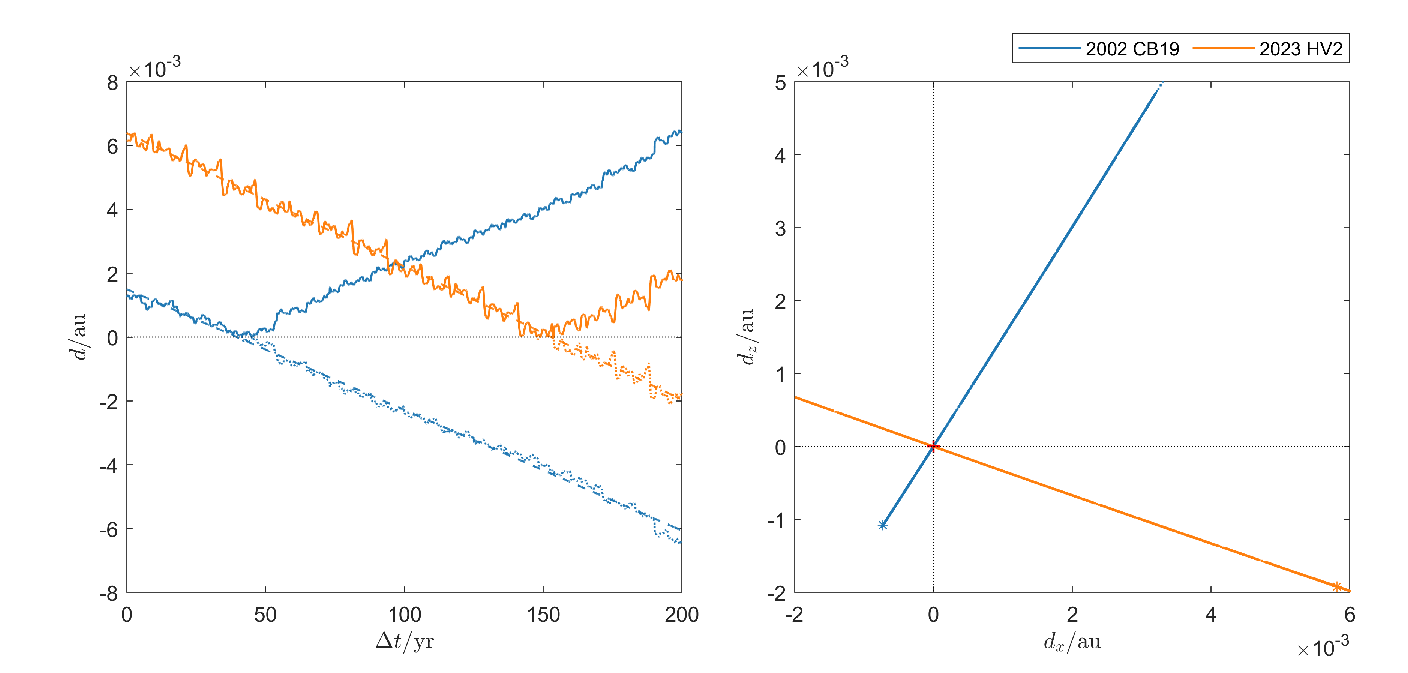}
    \caption{Fitting results of the MOID variations (left) and the corresponding critical points migration in the $x-z$ plane (right) for asteroids 2002 CB19 and 2023 HV2. In the left panel, the solid lines represent the precise values, while the dotted lines illustrate the signed MOID and the dashed lines demonstrate the fits of the signed MOID. The reference frame and symbols used in the right panel are identical to those employed in Fig. \ref{fig:fig_fit_good}.}
    \label{fig:fig_moid_cross}
\end{figure*}

Notably, the eight asteroids analyzed in the left panel of Fig. \ref{fig:fig_fit_good} exhibit no orbital crossings with the EMB over the 200-year period, preventing their MOID from approaching zero. In contrast, the left panel of Fig. \ref{fig:fig_moid_cross} illustrates two representative cases --- 2002 CB19 and 2023 HV2 --- which develop intersecting orbits with the EMB. The solid line represents the precise MOID, and an inflection point is observed during the orbital crossing. This feature stems from the mathematical singularity at $d = 0$, which precludes the linear fitting. To address this discontinuity, a sign inversion methodology is proposed during the orbital crossing events to preserve the analytical continuity of the derivative at the inflection point.

Previous investigations of signed MOID reveal distinct methodological approaches. \cite{valsecchi2003resonant} pioneered a sign convention where positive MOID values indicate critical points exterior to Earth's orbit, while negative values correspond to interior positions. Subsequent work by \cite{gronchi2013evolution} advanced this concept by constructing a signed distance function through vector orientation analysis at the local minima. This approach leverages the cross-product direction between orbital tangent vectors at local minimum points and its dot product with the relative position vector, creating a smooth analytic continuation across crossing configurations and effectively resolving the non-differentiability inherent in the classical distance function.

Our analysis demonstrates limitations in the former approach for objects exhibiting transitional crossings approximately along the direction perpendicular to the ecliptic. We therefore adopt the latter approach, initializing MOID as positive until orbital crossing events are confirmed. Under this framework, the fitted secular MOID drift $k$ carries physical significance: negative $k$ values correspond to decreasing Earth-approaching distances, while positive values indicate increasing separation.

The implementation of the signed MOID demonstrates significant improvement in linear fitting efficacy for orbit-crossing scenarios. As shown in the left panel of Fig. \ref{fig:fig_moid_cross}, the signed MOID (dotted lines, left panel) enables coherent tracking through orbital crossings for the two asteroids, with dashed lines representing optimized linear fits whose parameters are listed in Tab. \ref{tab:fits-example}. Our computational results indicate that 3,507 NEAs are predicted to experience the orbital crossings within the 200 years, confirming the broad applicability of this methodology in the linear fitting. Further analysis highlights a diverse distribution of MEI values among these asteroids. Specifically, the percentages of asteroids with MEI values ranging from 0.0 to 0.9 are 0.6\%, 2.4\%, 12.3\%, 24.2\%, 19.3\%, 14.0\%, 12.8\%, 9.2\%, 4.5\%, and 0.7\%, respectively.

\begin{table*}
    \centering
    \caption{Numerical characterization of MOID evolution for 14 selected near-Earth asteroids.}
      \begin{tabular}{clccccccccccc}
      \toprule
      NO. & Designation & $a$/au & $e$ & $i/^\circ$ & $\Omega/^\circ$ & $\omega/^\circ$ & ${{d}_0}$/au & $k$ & $\phi_0/^\circ$ & $\varepsilon$/au & $\eta$ & MEI \\
      \hline
      (1) & (2102) Tantalus & 1.290 & 0.299 & 64.0 & 94.4 & 61.5 & 0.0429 & -0.89 & 14.8 & 0.0002 & 0.01 & 5.4 \\
      (2) & (2201) Oljato & 2.179 & 0.711 & 2.5 & 74.9 & 98.4 & 0.0030 & 1.07 & 266.0 & 0.0012 & 0.15 & 1.4 \\
      (3) & (4183) Cuno & 1.981 & 0.636 & 6.7 & 294.4 & 237.0 & 0.0293 & -2.20 & 104.4 & 0.0019 & 0.10 & 3.5 \\
      (4) & (4953) 1990 MU & 1.621 & 0.657 & 24.4 & 77.5 & 77.9 & 0.0254 & -2.55 & 61.4 & 0.0008 & 0.06 & 2.6 \\
      (5) & (26663) 2000 XK47 & 1.546 & 0.472 & 13.5 & 303.6 & 231.2 & 0.0349 & -0.84 & 127.1 & 0.0007 & 0.02 & 5.4 \\
      (6) & (37655) Illapa & 1.478 & 0.752 & 18.0 & 139.5 & 303.9 & 0.0248 & 2.38 & 285.3 & 0.0005 & 0.01 & 4.5 \\
      (7) & (39572) 1993 DQ1 & 2.036 & 0.493 & 10.0 & 313.5 & 344.8 & 0.0350 & -0.72 & -- & 0.0077 & 0.24 & 4.5 \\
      (8) & (53429) 1999 TF5 & 2.022 & 0.641 & 26.8 & 199.2 & 64.1 & 0.0485 & -1.85 & 226.0 & 0.0046 & 0.11 & 5.5 \\
      (9) & 2002 CB19 & 1.283 & 0.452 & 17.1 & 318.0 & 88.8 & 0.0015 & -0.89 & 236.1 & 0.0005 & 0.14 & 0.4 \\
      (10) & 2023 HV2 & 1.616 & 0.423 & 44.0 & 32.4 & 222.4 & 0.0064 & -0.98 & 341.7 & 0.0007 & 0.16 & 0.4 \\
      (11) & 2018 GQ & 0.898 & 0.203 & 18.5 & 187.3 & 211.3 & 0.0029 & 15.91 & -- & 0.0616 & 0.68 & 5.8 \\
      (12) & 2006 DL & 2.501 & 0.738 & 4.8 & 323.1 & 94.9 & 0.0176 & -2.30 & 82.9 & 0.0081 & 0.68 & 0.5 \\
      (13) & 2002 CB26 & 1.950 & 0.724 & 6.8 & 139.5 & 265.9 & 0.0046 & 0.53 & 99.0 & 0.0050 & 0.81 & 0.5 \\
      (14) & 2001 FR85 & 0.983 & 0.028 & 5.2 & 183.0 & 233.6 & 0.0045 & -0.71 & -- & 0.0076 & 1.54 & 0.4 \\
      \hline
      \end{tabular}
      \begin{tablenotes}
        \item[1] Note: The orbital elements are the osculating values at the epoch $t_0$ of Jan 1, 2025. The unit of $k$ is $R_e$/yr.
    \end{tablenotes}
    \label{tab:fits-example}
\end{table*}

By employing the linear fitting to approximate MOID variations and implementing the sign inversion strategy for orbit-crossing cases, we are able to derive the three critical parameters: the initial MOID offset $d_0$, the linear rate coefficient $k$, and the uncertainty bound $\varepsilon$. The time-dependent MOID estimation can then be expressed as
\begin{equation}
    \label{eq:recovery}
    d'\left( t \right) = {d_0} + k\cdot\left( {t - {t_0}} \right) \pm \varepsilon
\end{equation}
where the modeling uncertainty $\varepsilon$ is also taken into account. Importantly, negative values of $d'$ are physically meaningless and should be replaced with its absolute value. If the upper or lower bound is considered and intersects the zero line, the value may need to be also replaced with 0. This parameterization enables efficient reconstruction of MOID evolutionary trends while maintaining awareness of the predictive limitations.

\begin{figure*}
    \centering
    \includegraphics[width=\textwidth]{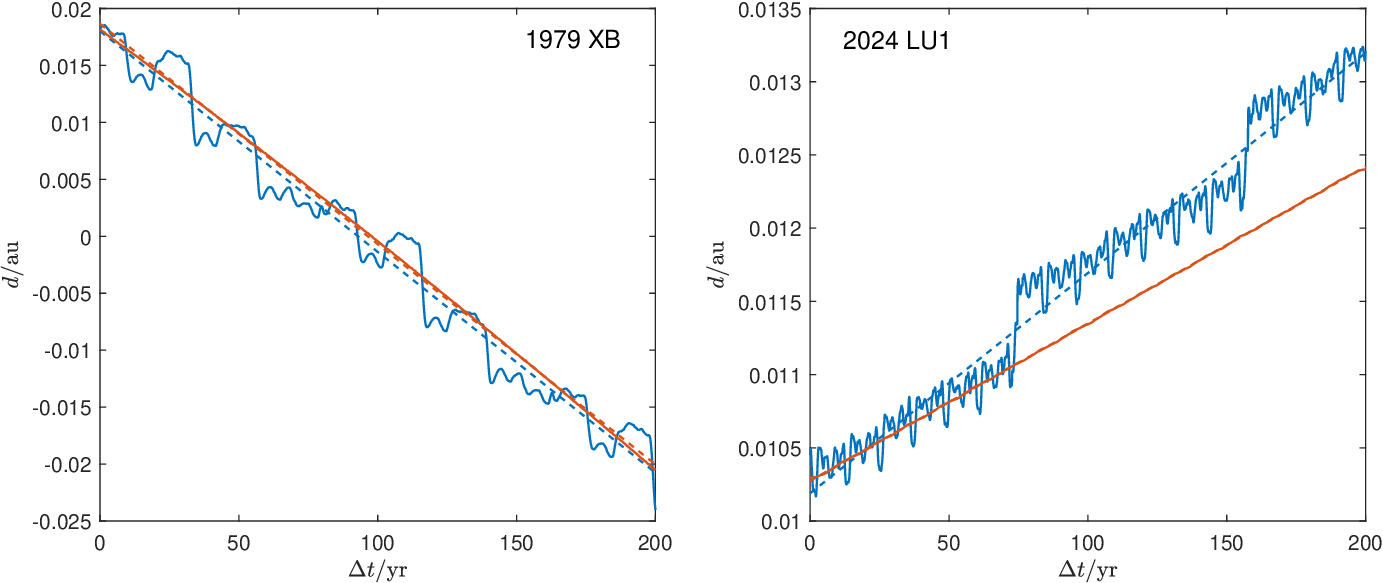}
    \caption{Comparative analysis of MOID fitting performance for two asteroids, 1979 XB (left panel) and 2024 LU1 (right panel), using the two different methods. The blue solid line represents the precise MOID values computed through orbit integration using full-force dynamical model, while the blue dashed line illustrates the fitting results obtained through our direct fitting approach. For comparison, the red solid line denotes the MOID values calculated using \cite{gronchi2013evolution}'s double average method, with the corresponding fitting results shown as red dashed lines.}
    \label{fig:fig_fit_bijiao}
\end{figure*}

Before we continue, it is important to note that \cite{gronchi2013evolution} have proposed a different methodology to obtain the secular MOID evolution. They computed the secular evolution of the orbital elements of the asteroid by using a double-averaged Hamiltonian framework to remove the mean anomalies of both the asteroid and planets, complemented by an innovative singularity extraction technique for resolving orbit-crossing configurations. For comparative analysis, Fig. \ref{fig:fig_fit_bijiao} illustrates the MOID evolution of asteroid 1979 XB and 2024 LU1, alongside the linear approximations. The blue solid lines represent the precise MOID values obtained from high-fidelity orbital integration, while the blue dashed lines denote the linear fits using our method. In parallel, the red solid lines show the MOID derived using \cite{gronchi2013evolution}'s approach, with the corresponding fits depicted as red dashed lines (since the MOID calculated using the double-averaged method has already eliminated the short-period variations, the red curve closely aligns with the fitted red dashed line). Both methods demonstrate comparable accuracy in capturing the secular linear trend for 1979 XB, with the maximum residual (the maximum difference with respect to the solid blue line) of 0.0036 au (our method) and 0.0043 au (their method), respectively. This quantitative agreement validates the effectiveness of our linear approximation for secular MOID evolution.

For 2024 LU1, however, our method achieves a maximum residual of 0.0004 au, significantly lower than the 0.001 au residual observed with their method. Note that the fitted red dashed line gradually deviates from the blue solid line for $\Delta t > 72$ yr, whereas our fit adheres to the solid line and achieves superior fitting performance over the entire curve. This discrepancy arises because 2024 LU1 undergoes close planetary encounters, which invalidate the fundamental assumption of their approach---namely, that the Delaunay elements $\sqrt{a}$ and $\sqrt{a(1-e^2)}\cos{i}$ remain constant over the asteroid's orbital evolution. These encounters induce non-negligible perturbations to the orbital elements, rendering their method's underlying approximations inadequate for accurately capturing the MOID dynamics. Moreover, since \cite{gronchi2013evolution}'s method uses a simplified force model (eg., the gravitational perturbations induced by the Earth-Moon system are treated as a unified entity), this approximation can lead to significant inaccuracies for low-MOID objects. Consequently, while the double-averaged approach provides valuable mathematical insight into the secular dynamics of MOID evolution, our direct fitting method offers a more straightforward and universally applicable solution. By circumventing the need for restrictive assumptions about orbital configurations, our approach maintains robustness across diverse dynamical regimes, particularly for objects experiencing frequent close encounters.

\subsection{Locations of critical points}
Critical points are defined as stationary points (including local minima, maxima, and saddle points) of the distance function between the asteroid and Earth's orbit. While the distance function may exhibit multiple local minima (each associated with a pair of critical points), only the global minimum defines the MOID \citep{gronchi2006mutual}. A systematic topological analysis of the critical points corresponding to the minimum of all local minima in the distance function can provide fundamental insights into the evolution of the MOID.

\begin{figure}
    \centering
    \includegraphics[width=0.47\textwidth]{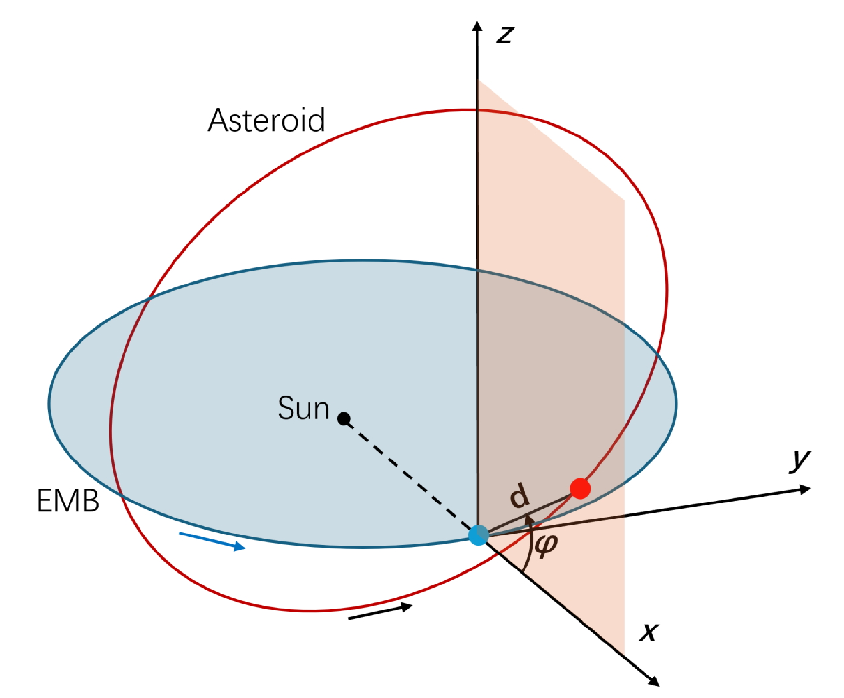}
    \caption{Illustration of the defined critical point migration frame. The light blue and red spheres represent the EMB (which is the origin of the coordinate system) and the asteroid position associated with the minimum distance $d$. In the case of a circular EMB orbit, the asteroid is situated within the $x-z$ plane (the angle $\phi$ is also illustrated).}
    \label{fig:fig_frame}
\end{figure}

\begin{figure*}
    \centering
    \includegraphics[width=\textwidth]{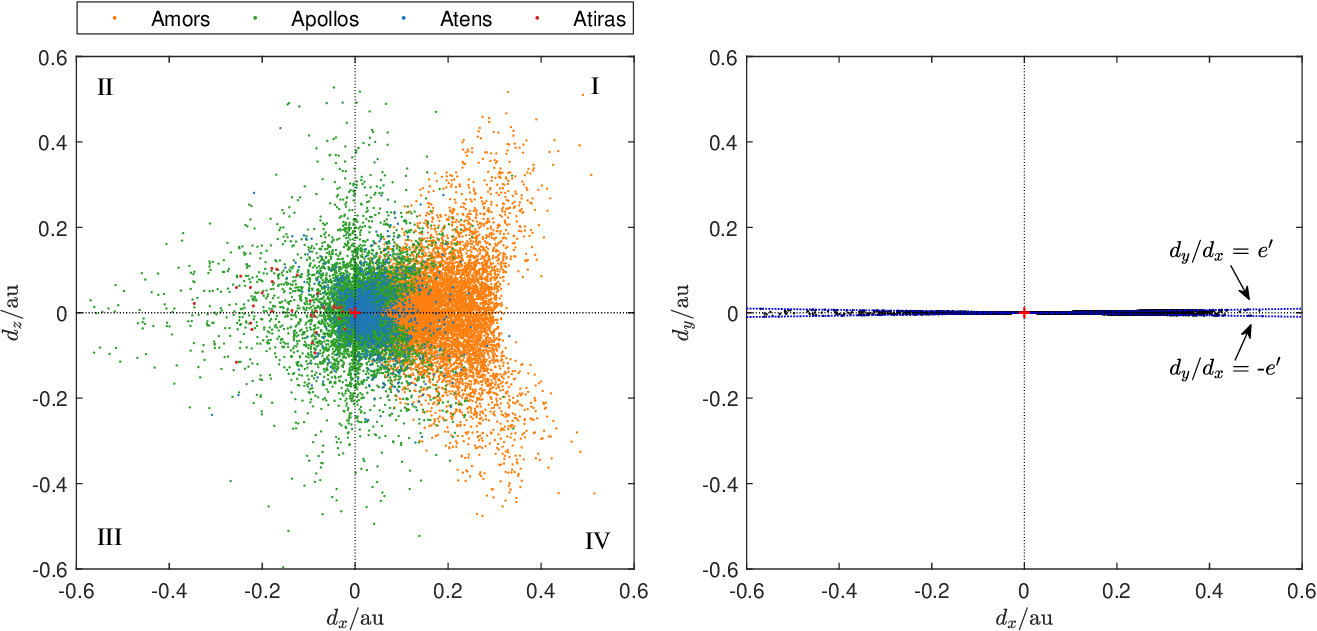}
    \caption{Distribution of the critical points at the initial epoch $t_0$ for the NEAs, projected in the $x-z$ (left, divided into four quadrants) and $x-y$ (right) planes of the migration frame, with the origin marked with a red ``+'' symbol (the correspondent critical points along the EMB orbit). In the left panel, the asteroids with different orbital types are represented by different colors. The two asymptotes described in the text, represented by the blue dotted line, are illustrated in the right panel.}
    \label{fig:fig_moid_pos_xz_xy}
\end{figure*}

Previous studies have indicated that the two critical points, in addition to the Sun, are situated in a plane perpendicular to the ecliptic plane under the assumption of a circular Earth orbit \citep{bonanno2000analytical, gronchi2013possible}. Although the EMB's actual eccentricity is $e' = 0.0167$, it is easy to prove that the plane containing these three points is still approximately perpendicular to the ecliptic plane. To better characterize the position of the critical points, we define a so-called critical point migration frame (which, to some extent, resembles the Sun-EMB co-rotating frame), wherein the origin is designated as the critical point on the EMB's orbit. The $x$-axis is established as the Sun-to-origin direction, the $z$-axis points toward the north ecliptic pole, and the $y$-axis is determined according to the right-hand rule (see Fig. \ref{fig:fig_frame}). Then the minimum distance vector ${\bf{d}}$ can be decomposed into three orthogonal components $(d_x, d_y, d_z)$, yielding $d = \sqrt{d_x^2+d_y^2+d_z^2}$. Each of the components carries a signed value determined by the positional vector of the critical point along the asteroid's orbit in the frame. Fig. \ref{fig:fig_moid_pos_xz_xy} shows the distribution of the critical points at the initial epoch $t_0$ for the NEA population, projected in the $x-z$ (left) and $x-y$ (right) planes of the migration frame. The left panel illustrates the broad distribution of critical points across four quadrants in the $x-z$ plane, while the right panel reveals that the points are confined within $d_y/d_x = \pm e'$, confirming the dominance of the $x$-$z$ plane in the spatial distribution. This feature allows us to approximately represent the position of the critical points via an azimuthal angle ($\phi$) in the $x-z$ plane,
\begin{equation}
    \label{eq:phi}
\phi  = \arctan \left( {\frac{d_z}{d_x}} \right),\;\;\phi  \in [0^\circ ,\;360^\circ )\;
\end{equation}
This parameter is consistent with the definition given by \cite{bonanno2000analytical}, which can be approximately expressed as
\begin{equation}
    \label{eq:phi_analytical}
    \frac{1}{{{{\cos }^2}\phi }} = 1 + \frac{{a\left( {1 - {e^2}} \right){{\sin }^2}i}}{{2 - 1/a - a\left( {1 - {e^2}} \right)}}
\end{equation}
The $\phi$ value is helpful to characterize the orientation of the asteroid's critical point relative to the EMB's in the migration frame. The $x-z$ plane is therefore divided into 4 quadrants (denoted as quadrant I, II, III, and IV) determined by the sign of $d_x$ and $d_z$. As shown in the left panel of Fig. \ref{fig:fig_moid_pos_xz_xy}, the critical point localization is related to the orbital type. That is, the Atiras are situated in the quadrants II and III ($d_x < 0$), whereas the Amors are located in the quadrants I and IV ($d_x > 0$). In contrast, the Atens or Apollos may be distributed over all the four quadrants, reflecting their orbital intersection dynamics.

Under the influence of orbital perturbations, the critical points exhibit systematic migration within the $x-z$ plane. Using the EMB-centric $d_B$ instead of the Earth-centric $d_E$ can enhance the regularity of the critical point migration trajectory, particularly for low-MOID cases, thereby facilitating a more comprehensive understanding of its secular MOID evolutionary behavior. The right panel of Fig. \ref{fig:fig_fit_good} demonstrates this dynamical evolution for the eight asteroids shown in the left panel, where asterisks denote the positions at $t_0$. The results reveal a distinct positive correlation between migration amplitude and MOID variation range, particularly evident in the contrasting behaviors of Tantalus and 1990 MU. Furthermore, asteroids exhibiting substantial MOID fluctuations (e.g., 1993 DQ1 and 1999 TF5) demonstrate correspondingly complex trajectory patterns on the plane.

\begin{figure}
    \centering
    \includegraphics[width=0.47\textwidth]{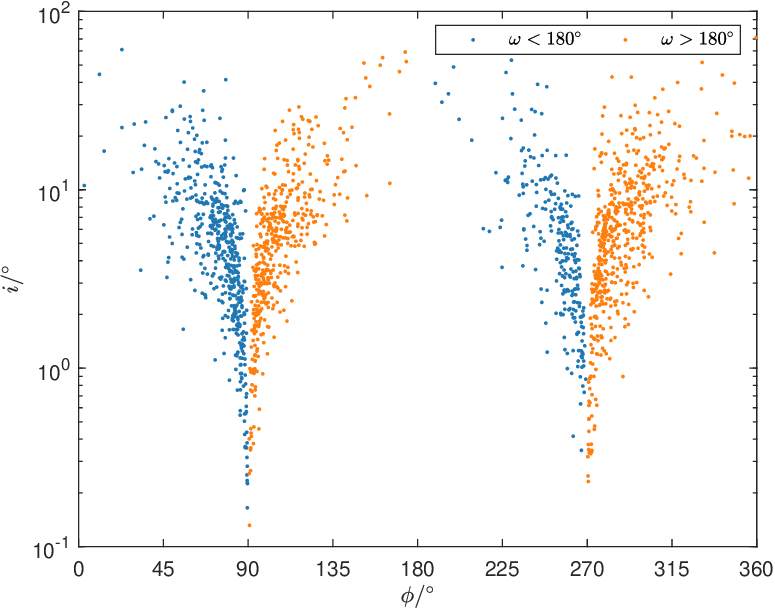}
    \caption{Distribution of $\phi$ versus $i$ for the four types of NEAs exhibiting a single crossing within the 200-year period. The parameter values are adopted at the initial epoch, with points in different ranges of $\omega$ highlighted in distinct colors.}
    \label{fig:fig_phi0_distribution}
\end{figure}

The two orbit-crossing cases presented in Fig. \ref{fig:fig_moid_cross} reveal distinct critical point migration patterns across the coordinate quadrants. Specifically, 2002 CB19 demonstrates a quadrant migration from III to I through the origin, while 2023 HV2 follows a path from IV to II. In general, for orbit-crossing cases, the trajectories of critical points toward the singularity in the $x-z$ plane can be categorized into four distinct types (${\phi _0}\equiv \phi \left( {{t_0}} \right)$):
\begin{itemize}
\item Type A: Migration from quadrant I to III ($\phi_0 \in [0^\circ, 90^\circ)$).
\item Type B: Migration from quadrant II to IV ($\phi_0 \in [90^\circ, 180^\circ)$).
\item Type C: Migration from quadrant III to I ($\phi_0 \in [180^\circ, 270^\circ)$).
\item Type D: Migration from quadrant IV to II ($\phi_0 \in [270^\circ, 360^\circ)$).
\end{itemize}
Here, type A and type D represent objects approaching Earth's orbit from the exterior to the interior direction, whereas type B and type C objects move from the interior to the exterior direction. Our statistical analysis shows that the migration patterns of 597, 502, 299 and 663 objects can be classified into type A, B, C, and D, respectively, among the 3,507 objects having orbital crossings. However, 1,066 objects exhibiting multiple crossings (see Fig. \ref{fig:fig_multiple} for details) and 380 objects demonstrating complex migration patterns cannot be classified into the four types. These complex patterns may involve migration between adjacent quadrants (e.g., from I to II, II to III, etc.) or can span across more than two quadrants, as exemplified by asteroid 2001 FR85 in the bottom panel of Fig. \ref{fig:fig_bad_fit}. This statistical analysis further underscores the intricate nature of MOID evolution for NEAs. Notably, the majority of currently known NEAs exhibit a preference for approaching Earth's orbit from the exterior direction. This trend may reflect the intrinsic orbit distribution of NEAs or, alternatively, could be attributed to observational selection effects. In fact, as demonstrated by Eq. \ref{eq:dmin2}, it can be readily shown that objects following type A and type C migrations predominantly have $\omega < 180^\circ$, while type B and type D objects tend to exhibit $\omega > 180^\circ$. This distinction is verified through Fig. \ref{fig:fig_phi0_distribution}, which illustrates the different distribution of $\phi$ for $\omega < 180^\circ$ and $\omega > 180^\circ$ among the 2,061 objects classified into the four categories.

Consequently, if only the three parameters $d_0$, $k$, and $\varepsilon$ are provided for two asteroids, we are unable to distinguish which path they take in the absence of the parameter $\phi$. This parameter is therefore essential for gaining deeper insights into how an asteroid's orbit approaches Earth's orbit. Specifically, as described by Eq. \ref{eq:phi_analytical}, asteroids with lower inclinations tend to display $\phi$ values concentrated around 90$^\circ$ or 270$^\circ$. This trend is further corroborated by Fig. \ref{fig:fig_phi0_distribution}, which illustrates the relationship between $\phi$ and $i$.

\begin{figure*}
    \centering
    \includegraphics[width=\textwidth]{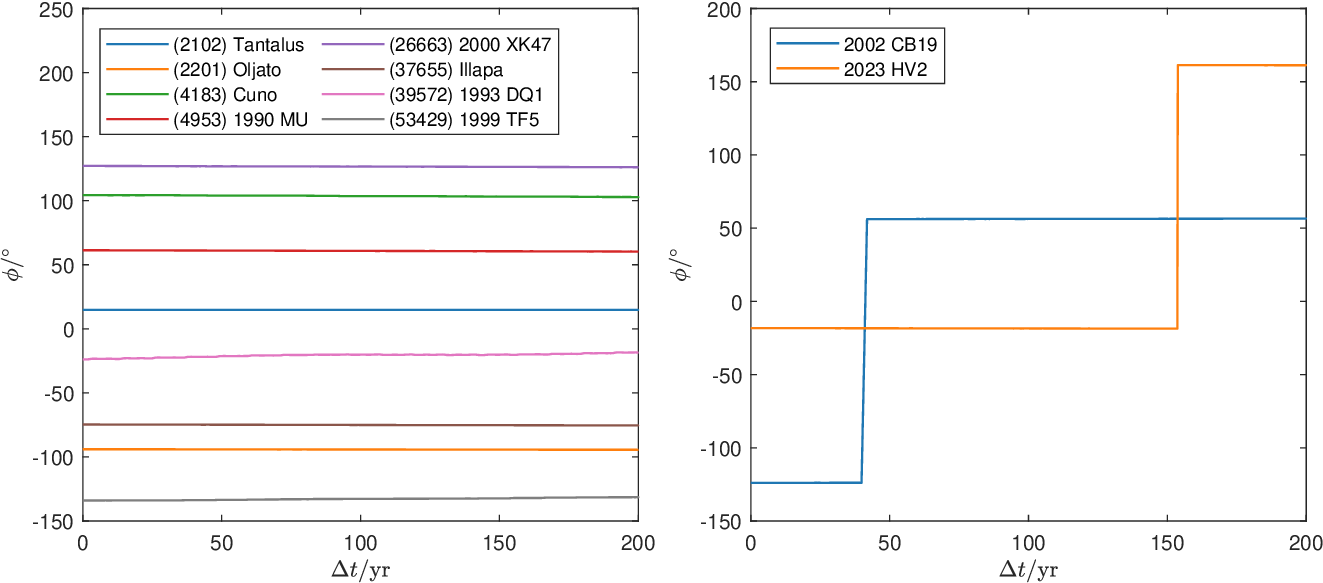}
    \caption{Variations of $\phi$ for the eight asteroids in Fig. \ref{fig:fig_fit_good} and the two asteroids in Fig. \ref{fig:fig_moid_cross}.}
    \label{fig:fig_phi_evol}
\end{figure*}

For the eight asteroids in Fig. \ref{fig:fig_fit_good} and the two asteroids in Fig. \ref{fig:fig_moid_cross}, the corresponding variations of $\phi$ are illustrated in the left and right panel of Fig. \ref{fig:fig_phi_evol}, respectively. We observe that the values of $\phi$ appear to exhibit only minor fluctuations for the eight non-crossing asteroids, in which only 1993 DQ1 exhibits a relatively large variation exceeding 5$^\circ$. The two orbit-crossing asteroids, however, exhibit distinct behavior through discrete 180$^\circ$ phase reversals at crossing epochs ($t_{\rm{cross}}$).

As indicated by the analytical solution Eq. \ref{eq:phi_analytical}, $\phi$ is independent of the argument of perihelion $\omega$, indicating that the secular perihelion precession (as well as the short-period fluctuation) does not fundamentally alter the orientation of the trajectory relative to the EMB on the $x-z$ migration plane. This result highlights the distinct roles of $\omega$ (driving temporal evolution) and $\phi$ (governing spatial orientation) in characterizing the MOID evolution, with $\phi$ instead being determined primarily by the coupled variations of $a$, $e$, and $i$.

We therefore propose to utilize solely the initial value $\phi_0$ to characterize the orientation of the critical points. However, for some asteroids situated within complex dynamical environments (close encounters, near mean-motion resonances, etc.), the amplitude of variations in $\phi$ can be very large. As precise values of $\phi$ are not typically required, we suggest that if this condition
\begin{equation}
    \label{eq:dphi_criterion}
    \left\{ \begin{array}{l}
    \max \left( {|\phi \left( {{t_i}} \right) - \phi_0|} \right) \le 5^\circ \;({t_i} < {t_{\rm{cross}}})\\
    \max \left( {|\phi \left( {{t_i}} \right) - \phi_0 + 180^\circ |} \right) \le 5^\circ \;({t_i} > {t_{\rm{cross}}})
    \end{array} \right.
\end{equation}
is satisfied, the $\phi_0$ parameter is used to characterize the orientation. If the condition is not met, $\phi_0$ is omitted, indicating that the migration direction exhibits significant variations. The results for selected asteroids are presented in Tab. \ref{tab:fits-example}. Statistical analysis reveals that 77\% NEAs in the catalogue exhibit a relatively stable $\phi$ value that satisfies Eq. \ref{eq:dphi_criterion}.

\subsection{MOID evolution characterization}
\label{section_MOID_characterization}
\begin{figure*}
    \centering
    \includegraphics[width=\textwidth]{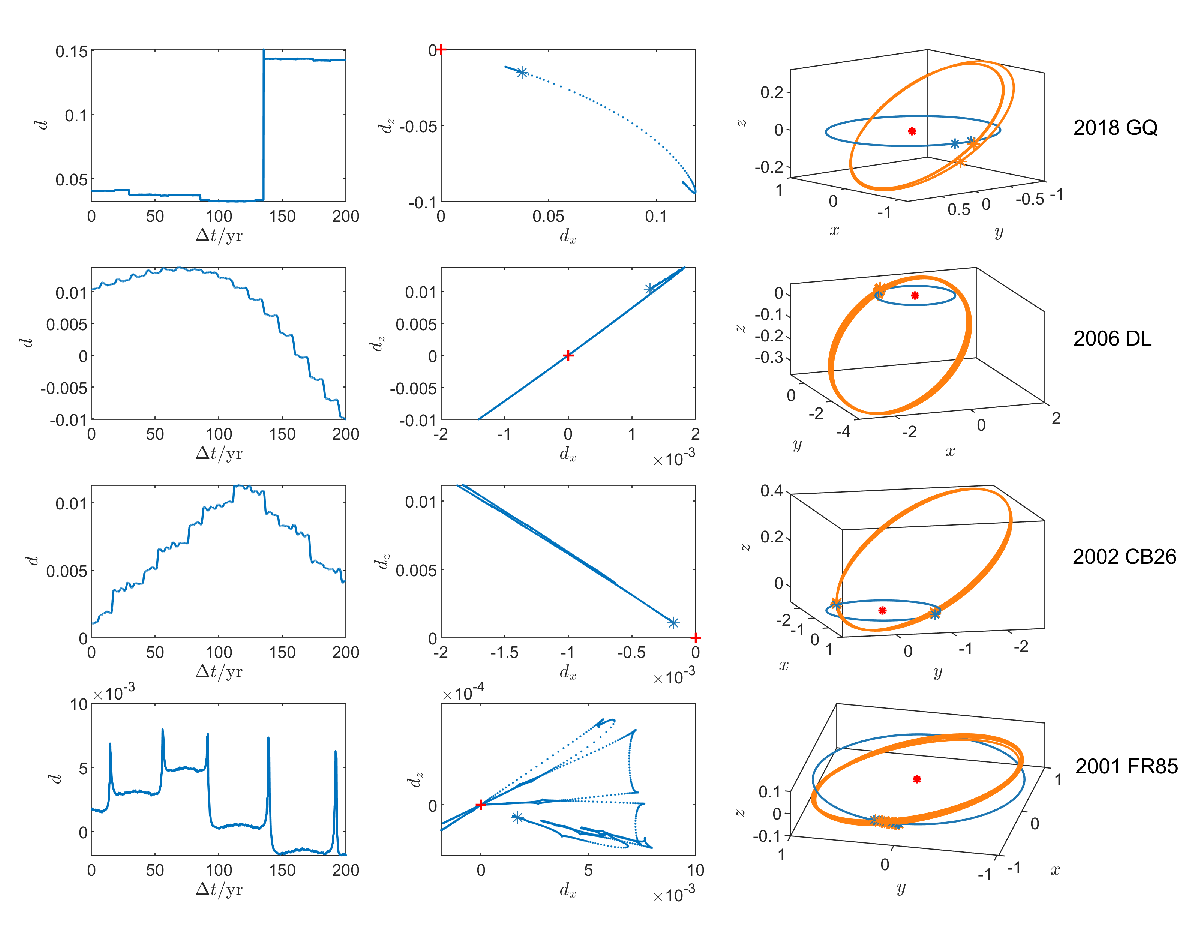}
    \caption{MOID propagations (panels in the left column), migration of the critical points (panels in the middle), and evolution of orbital configurations (panels in the right) for the four selected NEAs. For the case of 2006 DL and 2001 FR85, the MOID is assigned to be negative after orbit crossing. The asterisks in the middle panels indicate the initial position of the critical points, while the red plus symbol represents the origin of the frame. The orbits of the EMB and the asteroids are represented in blue and orange, respectively. Additionally, the critical points along both orbits are illustrated with asterisks. All the measurements of the MOID and orbit coordinates are in au.}
    \label{fig:fig_bad_fit}
\end{figure*}

The ten asteroids (1)-(10) in Tab. \ref{tab:fits-example} are the illustrative examples whose MOID evolution can be well approximated with the linear fit. However, some asteroids may exhibit distinctive orbital configurations or undergo significant perturbations, both of which can result in intricate MOID variations. As a result, the linear fit may prove to be an inadequate solution for such cases. The four asteroids 2018 GQ, 2006 DL, 2002 CB26, and 2001 FR85 listed in Tab. \ref{tab:fits-example} exemplify four distinct categories of exceptional cases that display such variations. The corresponding MOID evolution, migration of critical points projected in the $x-z$ plane of the critical point migration frame, and the orbital configuration evolution of the four cases are illustrated in the first, second, and third columns of Fig. \ref{fig:fig_bad_fit}, respectively. In what follows, we will discuss these objects in more detail.

(1) \textit{Close encounter}. As mentioned above, a close encounter with planet is an important mechanism to significantly change the orbit of an asteroid, and therefore change the MOID following the flyby. As shown in the top panels of Fig. \ref{fig:fig_bad_fit}, the asteroid 2018 GQ will undergo a close encounter with Venus in 2160 at a distance of 0.0003 au based on current orbit, leading to a change of 0.11 au in the MOID. Moreover, both the critical points have a large migration along the orbits. A large change of -$16.9^\circ$ is also observed in $\phi$ after the flyby (along with a rapid migration for the critical points in the plane), leading to the absence of $\phi_0$ in the table. The proportion of NEAs will experience close encounters with different planets is presented in Tab. \ref{tab:closeflyby}.

(2) \textit{Mean motion resonance}. Mean motion resonances (MMRs) with planets play a crucial role in the dynamical evolution of asteroids \citep{murray1999solar, morbidelli2002origin}. An asteroid in close proximity to a MMR can exhibit significant fluctuations in the orbital elements, resulting in a notable variation in the MOID. As an example, asteroid 2006 DL is in an approximate 3:1 resonance with Jupiter. The observed evolution of the MOID poorly follows a linear change, as can be seen in Fig. \ref{fig:fig_bad_fit}. MMRs with Jupiter are usually the dominant mechanisms to alter the orbit. However, MMRs with the terrestrial planets can be also prominent if the aphelion distance is low, otherwise the strong perturbation by Jupiter may diminish the dynamical behavior of the resonance \citep{de2013resonant}.

(3) \textit{Orbit with a special argument of periapsis}. Depending on the orbit geometry, the distance function may exhibit two local minima realized by two pairs of critical points close to the mutual orbital nodes \citep{gronchi2006mutual}. If the values of these two minima are close to each other, the MOID-associated critical points may undergo dynamical bifurcation---transitioning between the minima as the orbit evolves. As suggested by Eq. \ref{eq:dmin2}, if the argument of perihelion satisfies $\omega\approx90^\circ$ or 270$^\circ$, the two nodal distances may be approximately equal, resulting in a transition in the local critical points under minor perturbations. As shown by the results of 2002 CB26, whose $\omega$ is close to 270$^\circ$, there is an inflection point in the MOID. The transition cannot be observed in the $x-z$ plane, but can be clearly seen in the panel of orbital configuration evolution. If $\omega\approx0^\circ$ or 180$^\circ$, the inflection can also be observed in the MOID evolution, but the minima may not transition. In fact, for cases where $\omega\approx90^\circ$ or $270^\circ$, the MOID typically exhibits a non-differentiable rate of change. Conversely, for cases of $\omega\approx0^\circ$ or $180^\circ$, the MOID evolution remains differentiable at the inflection point.

(4) \textit{Orbit in 1:1 MMR with the Earth}. 
As a distinct subset of mean motion resonances, we classify Earth co-orbital objects (1:1 resonance) as a separate dynamical category. Their characteristic low relative velocities during Earth approaches enable prolonged gravitational interactions, leading to large variations in the orbital elements. Under these conditions, the analytical solution Eq. \ref{eq:dmin2} fails to characterize the MOID value and the solution Eq. \ref{eq:phi_analytical} fails to characterize the orientation angle $\phi$, particularly for low-eccentricity, low-inclination objects. The bottom panels of Fig. \ref{fig:fig_bad_fit} illustrate this phenomenon through asteroid 2001 FR85 ($a$ = 0.983 au, $e$ = 0.028, $i = 5.2^\circ$), which exhibits a substantial MOID shift of 0.001-0.002 au following an Earth flyby---a perturbation magnitude comparable to the initial MOID itself. Furthermore, the critical point trajectory displays complex migration pattern with multiple crossings in the plane. While dynamical studies \citep{milani1989dynamics,rabinowitz1994population,granvik2012population} suggest these objects remain impact-safe for several tens of thousands of years, their MOID values lose predictive reliability for long-term impact hazard assessment, rendering the detailed MOID evolution characterization redundant. Nevertheless, for database completeness, we retain the fits for these objects despite their limited physical interpretability. Notably, the presence of substantial short-period oscillations explains the relatively large $\Delta d_s$ at $a\approx1$ au observed in the right panel of Fig. \ref{fig:fig_MOID_pattern}.

\begin{figure*}
    \centering
    \includegraphics[width=\textwidth]{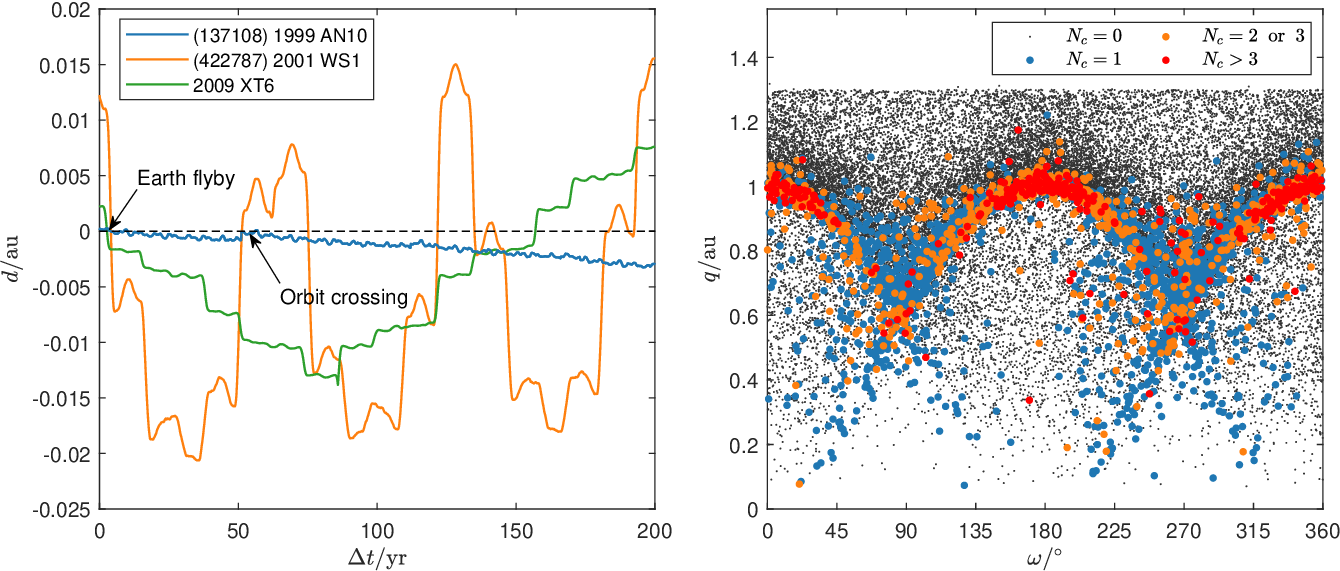}
    \caption{The left panel depicts the MOID evolution of three asteroids exhibiting multiple orbital crossings. The vertical coordinates are the signed MOIDs defined according to \cite{gronchi2013evolution}. The events of Earth flyby and orbit crossing are noted for 1999 AN10. The right panel illustrates the $\omega-q$ distribution (argument of perihelion vs. perihelion distance) for objects categorized according to their crossing numbers.}
    \label{fig:fig_multiple}
\end{figure*}

The above four categories are only a subset of cases that exhibit complex changes in the MOID. Nevertheless, combinations of these effects are also conceivable. For instance, depending on the orbital configuration and the dynamical environment, a close encounter might be accompanied by an MMR, or an MMR could be coupled with a local MOID transition. It is worth noting that certain objects may experience multiple orbit crossings. To exemplify this, the MOID propagations for three NEAs, namely (137108) 1999 AN10, (422787) 2001 WS1, and 2009 XT6, are presented in the left panel of Fig. \ref{fig:fig_multiple}. For 1999 AN10, a close flyby with Earth at $\Delta t$ = 4 years could result in a second crossing occuring at $\Delta t$ = 51 years subsequent to the initial one. The asteroid 2001 WS1 has a significant aphelion distance of 4.2 au. This implies that the perturbation exerted by Jupiter may cause a substantial change in the asteroid's orbital elements after it passes its aphelion. As can be seen in the figure, over the course of 200-year evolution, more than eight crossings are detected. Regarding the asteroid 2009 XT6, its argument of periapsis is approximately 90$^\circ$. This leads to a local MOID transition at $\Delta t$ = 80 years, which in turn results in a second crossing at $\Delta t$ = 156 years.

To better understand the population of near-Earth asteroids exhibiting multiple orbital crossings during the 200-year period, the right panel of Fig.\ref{fig:fig_multiple} displays the distribution of the argument of perihelion ($\omega$) versus perihelion distance ($q$) for objects with varying crossing numbers ($N_c$). We find that the single-crossing objects cluster near $\omega = 90^\circ$ or $270^\circ$ and are predominantly classified as Apollos or Atens, while multi-crossing objects concentrate near $\omega = 0^\circ$ or $180^\circ$, with orbital parameters closely resembling those of Amors ($q \approx 1$ au). For the latter group, when $\omega$ approaches $0^\circ$ or $180^\circ$, the average nodal distance (see Eq. \ref{eq:x0}) exhibits prolonged stability with minimal temporal variation. This dynamical behavior, combined with gravitational perturbations from Jupiter, enables repeated orbital intersections over extended timescales.

The scenarios discussed above inevitably generate complex nonlinear dynamics in the MOID evolution, fundamentally invalidating the linear approximation framework. To systematically evaluate the model adequacy, we establish a normalized goodness-of-fit metric $\eta$ through conditional definitions:
\begin{equation}
    \label{eq:eta}
    \eta  = \frac{\varepsilon }{{\bar d}},\;\bar d = \;\left\{ \begin{array}{l}
\left( {{d_1} + {d_2}} \right)/2\;\;\;\;\left( {{\rm{No}}\;{\rm{orbit}}\;{\rm{crossing}}} \right)\\
\left( {{{\tilde d}_2} - {{\tilde d}_1}} \right)/2\;\;\;\left( {{\rm{Orbit}}\;{\rm{crossing}}\;{\rm{present}}} \right)
\end{array} \right.
\end{equation}
where ${{{\tilde d}_1}}$ (negative) and ${{{\tilde d}_2}}$ (positive) denote the minimum and maximum signed MOID within the specified timespan, respectively.

\begin{figure*}
    \centering
    \includegraphics[width=\textwidth]{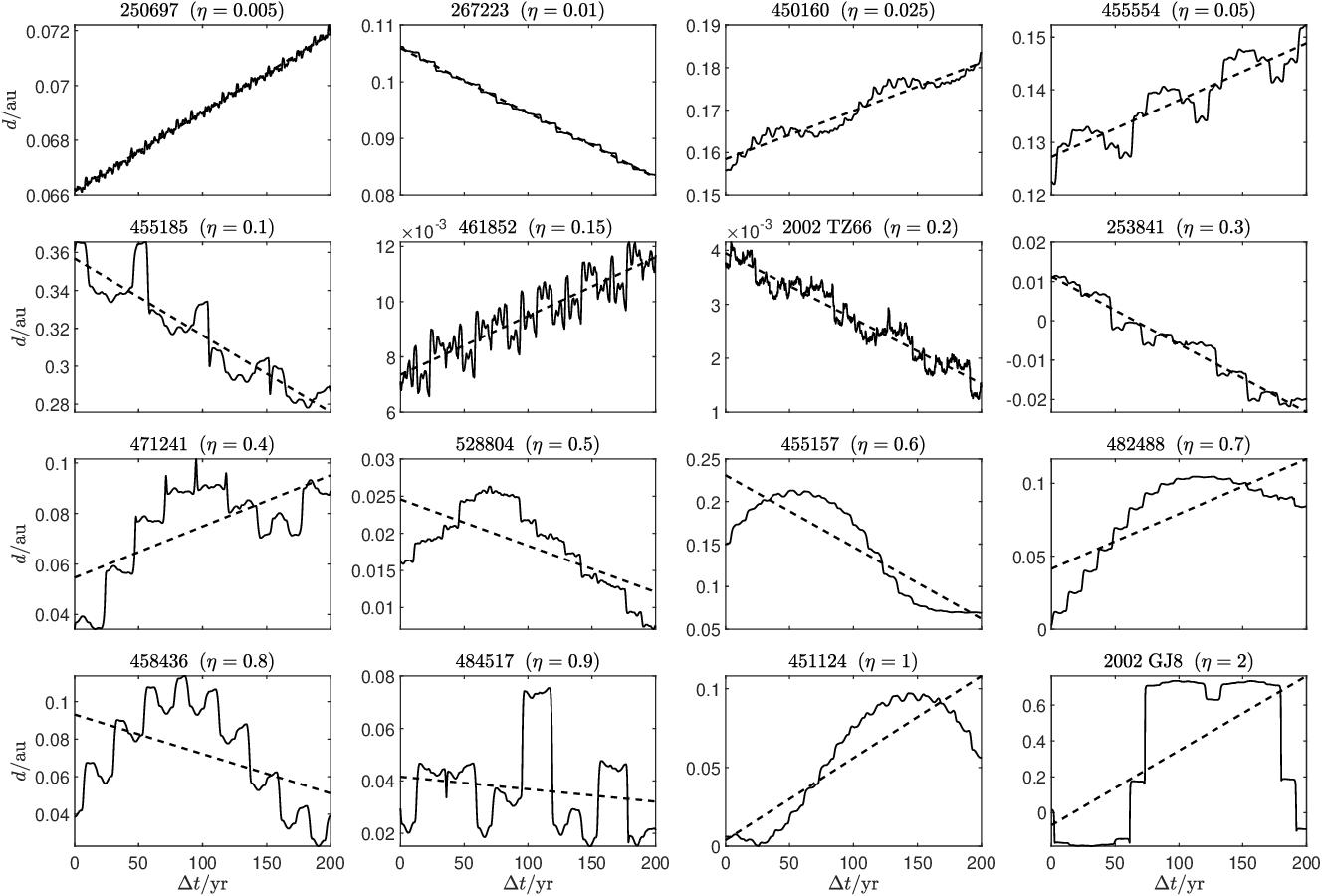}
    \caption{Comparison of MOID evolution (solid curves) and the linear approximations (dashed lines) for 16 selected NEAs spanning $\eta$ values from 0.005 to 2. The sign inversion strategy is applied to asteroids 253841, 451124, and 2002 GJ8, yet reveals significant discrepancies ($\eta > 0.3$) in the latter two cases.}
    \label{fig:fig_fit_diff_eta}
\end{figure*}

To establish a reliability criterion for our model, Fig. \ref{fig:fig_fit_diff_eta} shows the fitting results of 16 selected NEAs with varying magnitudes of $\eta$. A quantitative analysis reveals that about $70\%$, $81\%$, $87\%$, $90\%$, and $92\%$ NEAs satisfy $\eta \le 0.1$,  $\eta \le 0.2$, $\eta \le 0.3$, $\eta \le 0.4$, and $\eta \le 0.5$, respectively. Based on the results in the panels, we empirically define
\begin{equation}
    \label{eq:eta_threshold}
    \eta \le 0.3
\end{equation}
as the reliability threshold for an acceptable linear approximation---a criterion that excludes 13\% of NEAs in the catalogue. Notably, under this definition, a fit with a large $\varepsilon$ and a significantly larger $\bar d$ (eg., $\varepsilon$ = 0.01 au and $\bar d$ = 0.1 au) could still be deemed accepted. Comprehensive statistical details characterizing the $\eta$-distribution across the entire NEA population are provided in Section \ref{statistical_analysis}.

In summary, the following six parameters were adopted for each NEA to characterize the behavior of MOID evolution:
\begin{itemize}
    \item $d_0$: MOID value on the linear curve at initial epoch $t_0$.
    \item $k$: Secular drift rate obtained from the linear fit, mainly driven by the periapsis precession.
    \item $\phi_0$: An angle characterizing the orientation of the critical point, defined at $t_0$, which is only retained for objects that satisfy the criterion in Eq. \ref{eq:dphi_criterion}.
    \item $\varepsilon$: The peak residual deviation, quantifying the maximum deviation from linearity.
    \item $\eta$: A parameter used to characterize the relative peak residual deviation from linearity, which is also used to judge the goodness-of-fit.
    \item MEI: A defined MOID evolution index to quickly assess the MOID variations.
\end{itemize}

Consequently, we established a low-storage MOID evolution characterization database for the entire NEA population \footnote{For comprehensive details, Please refer to this GitHub resource: https://github.com/runningabc/MOID/}. This framework enables rapid assessment of MOID temporal variations and identification of potential Earth-approaching objects. The linear fitting parameters facilitate computationally efficient reconstruction of MOID values at arbitrary epochs within the 200 years, while the compact storage ensures scalability for future large-scale NEA population screening.

To demonstrate the operational efficacy, we employ this database to screen NEAs whose MOID may decline below a critical threshold $d_c$ (e.g., one lunar distance) within the 200-year period through a two-stage screening framework:

\textit{Step 1: MEI-preselecting}

First, we select NEAs with MEI $<$ 2.0 (here the cases of $m = 0, 1$ and $n = 0, ..., 9$ are included) to quickly exclude objects with a large MOID exceeding the threshold, which serves as a conservative filter to reduce the initial candidate pool.

\textit{Step 2: MOID evolution filtering}

Then, from the MEI-preselected subset, we identify objects satisfying either:
$$
d_0 - \varepsilon < d_c
$$
or
$$
d_0 + 200k - \varepsilon < d_c
$$
where the three fitting parameters $d_0$, $k$, and $\varepsilon$ are used.

This hierarchical screening identifies 6,053 candidate NEAs, compared to 5,789 confirmed through direct numerical verification---a 4.6\% false-positive rate that achieves an 83\% reduction in computational cost relative to full N-body simulations.

Note that the MEI consists of two digits $m$ and $n$, a simple filter (e.g., MEI < threshold) may not be sufficiently effective sometimes, and we may need to filter the $m$ and $n$ value separately.

\subsection{Statistical analysis}
\label{statistical_analysis}
\begin{figure*}
    \centering
    \includegraphics[width=\textwidth]{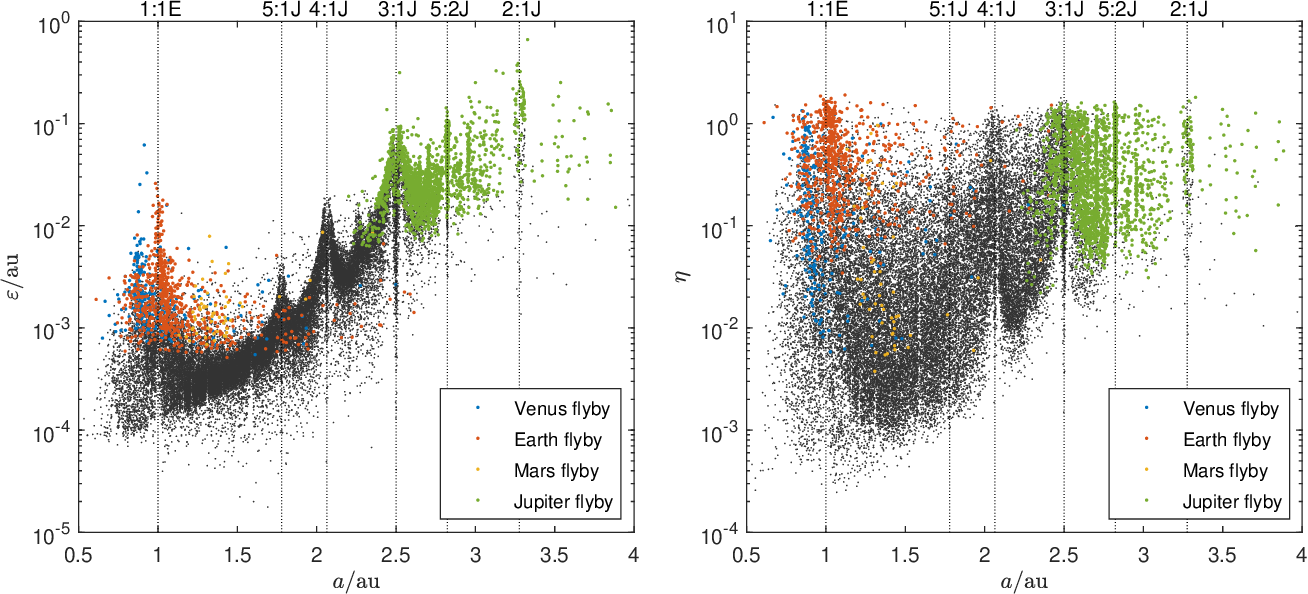}
    \caption{Distributions of $\varepsilon$ and $\eta$ as a function of the semi-major axis $a$. Objects that experience close encounters with different planets are represented by different colors. The locations of several MMRs are also illustrated for reference.}
    \label{fig:fig_epsl_eta}
\end{figure*}

The multi-parameter MOID evolution characterization database facilitates systematic investigation of parameter correlations under diverse orbital configurations and perturbative environments, which is helpful to gain insight into the MOID evolution dynamics. Crucially, our analysis reveals order-of-magnitude variations in the prediction accuracy across individual NEAs. Fig. \ref{fig:fig_epsl_eta} presents these variations through $\varepsilon$ and $\eta$ distributions versus semi-major axis $a$, revealing distinct dynamical regimes:

\begin{itemize}
    \item The $\varepsilon$ distribution exhibits a strong correlation with $a$, principally governed by Jupiter's gravitational perturbation beyond 2.0 au.
    \item Mean-motion resonances induce significant $\varepsilon$ increases, particularly in 1:1E, 5:1J, and 4:1J configurations where $\varepsilon$ values typically remain below 0.01 au, contrasting with 0.1 au deviations observed in other resonant populations (3:1J, 5:2J, 2:1J).
    \item Planetary close encounters (color-coded) further degrade the model accuracy, with Earth-approaching objects exhibiting particularly complex dynamics when coupling with the 1:1E MMR, which amplifies the nonlinear effects in their MOID evolution.
\end{itemize}

Overall, for objects with $a < 2$ au, 80\% achieve high precision ($\varepsilon < 0.001$ au), while it is only 3\% for objects with $a > 2$ au. In contrast, the normalized uncertainty $\eta$ shows reduced orbital dependence due to its intrinsic scaling with $\bar d$. This allows for acceptable linear approximations in 94\% of NEAs with $a < 2$ au and 73\% of those with $a > 2$ au, including the prevalent Amor objects that may have elevated $\bar d$ values to reduce the $\eta$ values.

\begin{figure*}
    \centering
    \includegraphics[width=\textwidth]{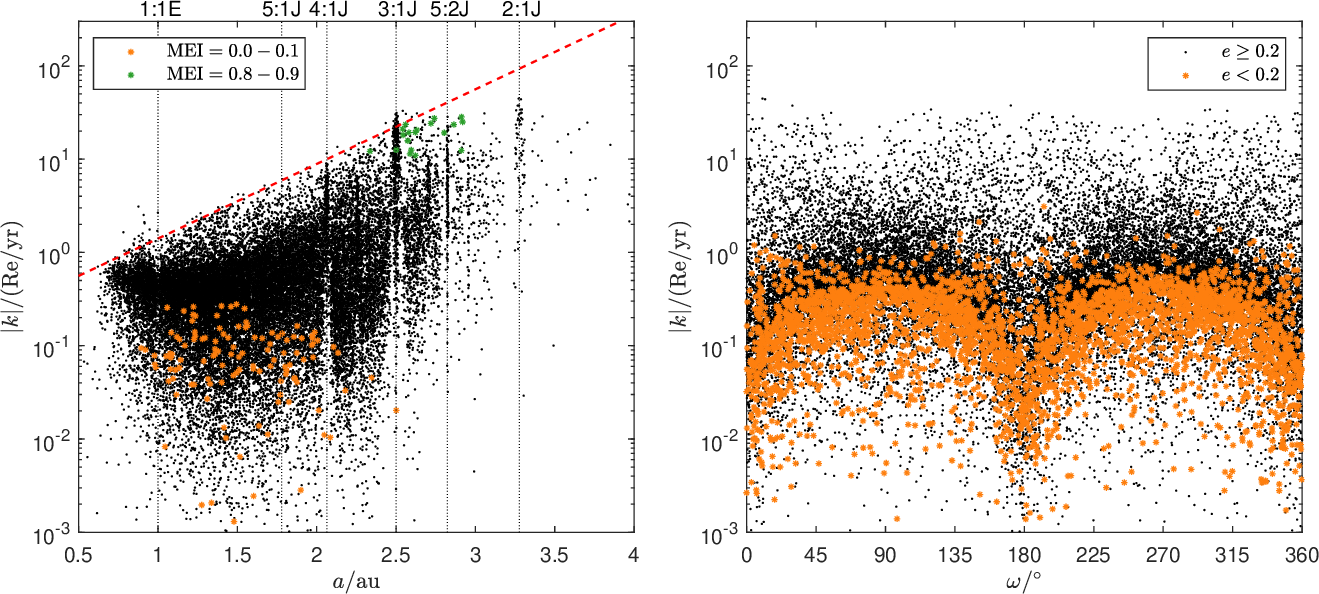}
    \caption{Distributions of secular drift rate $|k|$ as a function of $a$ (left panel) and $\omega$ (right panel) for NEAs satisfying Eq. \ref{eq:eta_threshold}. Objects with distinct MEI values (0.0--0.1 and 0.8--0.9) and eccentricity ranges ($e \ge 0.2$ or $e < 0.2$) are highlighted with asterisks in the panels. The red dashed line in the left panel represents the empirical upper limit defined by Eq. \ref{eq:upper_rate}}
    \label{fig:fig_rate_w}
\end{figure*}

As shown in Fig. \ref{fig:fig_rate_w}, the secular drift rate $|k|$ (absolute value) is plotted as a function of semi-major axis $a$ and argument of perihelion $\omega$ (only the asteroids satisfying the threshold Eq. \ref{eq:eta_threshold} is plotted). A pronounced correlation emerges between $|k|$ and $a$: larger semi-major axes correspond to systematically higher drift rates (left panel), which is also primarily driven by Jupiter's perturbation. This trend provides critical context for interpreting different distribution of $a$ among objects with MEI values of 0.0-0.1 versus 0.8-0.9 (asterisks in left panel). The MEI = 0.0-0.1 population exhibits low drift rates ($|k| < $0.3 $R_e/\rm{yr}$) and preferentially occupies the low-$a$ regime ($a < 2$ au). An empirical upper bound for $|k|$ (red dashed curve, left panel) is formulated as
\begin{equation}
    \label{eq:upper_rate}
    |k| < 0.6 \times {10^{ - 4}} \times {10^{0.8\left( {a - 1} \right)}}\;\left({\rm{au/yr}}\right)
\end{equation}
or
\begin{equation}
    \label{eq:upper_rate_Re}
    |k| < 1.4 \times {10^{0.8\left( {a - 1} \right)}}\;\left( {{R_e}{\rm{/yr}}} \right)
\end{equation}

Furthermore, the distribution in the right panel of Fig. \ref{fig:fig_rate_w} shows that the rate also strongly correlates with the argument of perihelion $\omega$. Specifically, low-drift objects cluster near $\omega=0^\circ$ and $180^\circ$, while high-drift populations dominate near $\omega=90^\circ$ and $270^\circ$. When $\omega$ approaches $\omega=0^\circ$ or $180^\circ$, the asteroid's perihelion/aphelion aligns with its orbital node. This configuration causes the projection of the perihelion/aphelion's tangential direction on the ecliptic plane to align nearly parallel with Earth's orbital motion. Consequently, the precession of $\omega$ exerts minimal influence on the nodal distance, thereby minimizing the secular drift of the MOID. Additionally, low orbital eccentricity also suppresses the drift by reducing the drift of the nodal distance: objects with $e<0.2$ typically have a rate less than 1 $R_e/\rm{yr}$. These phenomena could be also corroborated by the analytical solution derived in Eq. \ref{eq:dmin2}.

\section{Summary and Conclusions}
NEAs represent a class of small bodies that may, under certain circumstances, pose a potential threat to the safety of Earth. While the MOID serves as a fundamental parameter for impact risk assessment, its dynamic evolution under complex gravitational interactions necessitates systematic investigation. This study advances our understanding through a comprehensive 200-year dynamical analysis of all cataloged NEAs, introducing an innovative classification methodology and a numerical characterization framework. Our principal contributions are summarized as follows:

\begin{enumerate}
    \item We conducted a comprehensive quantitative analysis of the short-period perturbations induced by the Moon and the secular perturbations caused by other planets on Earth's MOID. Our findings reveal that lunar perturbation can induce periodic variations of up to 0.002 au, while planetary perturbations can drive secular drifts of up to $0.06\,R_e/\text{yr}$. Through the analysis of nine post-discovery impactors, we demonstrated that the EMB-centric MOID offers a superior predictive capability compared to the Earth-centric MOID, though the latter one is more critical for imminent (within a few days) impact scenarios. For the purposes of this paper, the EMB-centric MOID is adopted.
    \item To enable rapid assessment of MOID evolution, we introduced the MOID Evolution Index (MEI), a two-digit metric denoted as $m.n$, where values range from $0.0$ to $9.9$. This index is derived from the maximum and minimum MOID values over a $200$-year period: the first digit ($m$) quantifies the minimum value, while the second digit ($n$) captures the maximum variation range. The MEI serves as an intuitive and efficient tool for prioritizing NEAs for further investigation and identifying accessible targets for planetary science missions, with lower MEI values generally correlate with higher-risk priority and smaller minimum geocentric distances (in fact, the MOID provides the lower limit).
    \item We employed a linear function to model the MOID variations of all cataloged NEAs, incorporating a sign inversion strategy to address orbit-crossing cases. Based on our defined criteria, $87\%$ of NEAs were found to be well-approximated by the linear model. Furthermore, we proposed a frame to visualize the migration of the critical point in the plane. By defining an angle to characterize the orientation of the critical point relative to the EMB's orbit, we identified four distinct migration categories for orbit-crossing objects, providing new insights into their dynamical behaviors.
    \item We established a comprehensive database incorporating the MOID fitting parameters and MEI values, confirming its efficacy as a robust methodology for screening NEAs. Statistical analysis revealed correlations between the fitting parameters and orbital elements, offering valuable insights into the MOID evolution dynamics. From these results, we derived the upper bounds of the secular MOID drift rate as a function of semi-major axis, providing an important reference for long-term hazard assessment.
    \item Our analysis confirmed that certain special orbital configurations and complex dynamical environments can reduce the feasibility of linear fitting for some objects. In particular, we examined a unique class of orbits characterized by multiple orbit crossings. These findings underscore the limitations of relying on a fixed MOID value at a given time to assess the safety of NEAs, highlighting the intricate MOID evolution patterns for some NEAs.
\end{enumerate}

Finally, an important caveat is that this work does not account for the MOID uncertainties arising from the orbital element errors. Prior studies have addressed such uncertainties in cases of non-positive definiteness in covariance matrices \citep{bonanno2000analytical,gronchi2006mutual,gronchi2007uncertainty}. The theoretical framework would involve integrating a similar methodology to incorporate additional uncertainty terms into the calculation of $\varepsilon$. However, applying this methodology to the entire near-Earth asteroid population may pose challenges due to the heterogeneity of their orbital configurations and covariance characteristics, which will be investigated in future work. However, since the MOID propagation exhibits relative insensitivity to errors in mean anomaly (which is the dominant source of orbital uncertainty), the fitting parameters for many objects likely retain validity without the need to update frequent if the orbit precision achives some threshold. Thus, orbital uncertainties may not constitute a critical limitation in MOID evolution analysis for the these cases.

\begin{acknowledgments}
    We would like to express our sincere gratitude to both reviewers for their time and insightful comments. This research was supported by the Special funding project for space debris and near-Earth Asteroids defense research (KJSP2023020303).
\end{acknowledgments}

%%\clearpage
\bibliography{ms}{}
\bibliographystyle{aasjournal}

\end{document}